%% file: paper.tex
\documentclass{iopjournal}
\usepackage[numbers,sort&compress]{natbib}
\usepackage{upgreek}
\usepackage{amsmath}
\usepackage{amssymb}
\usepackage{tikz}
\usetikzlibrary{positioning,arrows.meta,calc}
\usepackage{bm}        

\begin{document}


\title{Physics-informed Bayesian optimization of high-order harmonic generation in gases}

\author{Dominika Maslarova$^{1,*}$\orcid{0000-0002-2756-3722},
Mattys Pouyez$^1$\orcid{0009-0005-9057-3228},
Patrik Jansson$^2$\orcid{0000-0003-3078-1437},
Jan Vabek$^{3,4}$\orcid{0000-0001-7689-7174}, \\
Robin Weissenbilder$^5$\orcid{0000-0002-8081-3727},
Linus Hagedorn$^{1,3}$\orcid{0009-0002-7207-9080},
Anne L'Huillier$^5$\orcid{0000-0002-1335-4022}, and
Tünde Fülöp$^{1}$\orcid{0000-0002-5898-0393}}

\affil{$^1$Department of Physics and Astronomy, Chalmers University of Technology, Gothenburg, Sweden}

\affil{$^2$Department of Computer Science and Engineering, Chalmers University of Technology and University of Gothenburg, Gothenburg, Sweden}

\affil{$^3$Czech Technical University in Prague, Faculty of Nuclear Sciences and Physical
Engineering, Prague, Czech Republic}

\affil{$^4$ELI Beamlines Facility, The Extreme Light Infrastructure ERIC, Dolní
Břežany, Czech Republic}

\affil{$^5$Department of Physics, Lund University, Lund, Sweden}

\email{$^{*}$dommas@chalmers.se}


\begin{abstract}
Optimization of high-order harmonic generation in gases is essential for attosecond science.
Improving conversion efficiency from the infrared driving laser to the extreme-ultraviolet high-order harmonic radiation requires balancing microscopic single-atom dipole responses with macroscopic phase matching.
The resulting conversion-efficiency landscape exhibits complex structures, such as the hyperbolic ridge in the pressure--length space, which can be difficult to accurately reconstruct using conventional Bayesian optimization.
Here, we introduce a physics-informed Bayesian optimization framework that incorporates known physical scaling relations through nonlinear transformations of the search space.
We first demonstrate the approach in the two-dimensional pressure--length space, where a direct grid scan enables detailed validation of the reconstructed landscape, before extending it to four dimensions, where exhaustive scanning becomes computationally impractical.
In four dimensions, incorporating the pulse-duration scaling and a square-root transformation of conversion efficiency further improves the accuracy of the reconstructed landscape,
as validated against an independent reference dataset. 
We also employ a threshold-based acquisition strategy to identify extended high-efficiency operating regions rather than isolated optimal points.
Together,
these results demonstrate that incorporating physical knowledge into Bayesian optimization improves the
exploration and reconstruction of multidimensional high-order harmonic generation landscapes, extending
its capabilities beyond locating individual optima.
\end{abstract}

\section{Introduction}
High-order harmonic generation (HHG) is a widely used method for producing coherent, short-pulse extreme-ultraviolet (XUV) radiation for fundamental studies of ultrafast electron dynamics \cite{KrauszIvanov,goulielmakis2010real} and for applications such as coherent imaging \cite{zurch2014real} or coincidence spectroscopy \cite{mikaelsson2020high}.
In a typical gas-based HHG experiment \cite{mcpherson1987studies,ferray1988multiple, brabec2000intense}, an intense femtosecond infrared (IR) laser is focused into a gaseous target. 
This  leads to the generation of a broad comb of discrete peaks at odd multiples of the laser frequency, reaching across the XUV range and potentially extending even into the soft-X-ray regime \cite{sheehy1999high,popmintchev2012bright}.
In the time domain, the coherent superposition of high harmonics forms an attosecond pulse train \cite{paul2001observation,agostini2004physics}.
The optimal operation of HHG sources is essential for addressing future challenges in chirality, quantum optics, and spin dynamics~\cite{attosecond_roadmap}, with a recent perspective highlighting broader developments in HHG sources and their applications to quantum phenomena \cite{Constant2025perspective}.
This requires advanced optimization techniques, as HHG is inherently a multiscale problem~\cite{5D_modelling_roadmap}, involving both the microscopic response of individual atoms to the driving laser field and the macroscopic propagation and buildup of the generated harmonic radiation. 
The interplay between these processes determines the overall IR-to-XUV conversion efficiency (further referred to only as conversion efficiency), which remains relatively low, typically on the order of $10^{-7}$ to $10^{-4}$~\cite{tempea2000optimization, brichta2009comparison, takahashi2013attosecond, nayak2018multiple, bergues2018tabletop, popmintchev2018near, weissenbilder2022optimize}.

Microscopically, HHG originates from the nonlinear response of individual atoms to the strong driving field. 
This interaction is described quantum mechanically by the time-dependent Schrödinger equation (TDSE) \cite{carlstrom2022generala,carlstrom2022generalb, vabek2026modularmultiscaleapproachmodelling}. Its essential physics can, however, be captured by the simplified semiclassical three-step model \cite{Corkum,Lewenstein}, which describes HHG in terms of tunnel ionization, acceleration in the continuum, and recombination.
First, the intense laser field distorts the atomic potential sufficiently for an electron to tunnel ionize. 
The electron is then accelerated by the field and gains kinetic energy. 
When the oscillating field reverses, the electron can be driven back toward its parent ion, recombine, and release its accumulated energy as a high-energy photon.
Repetition of this process over successive half-cycles gives rise to the characteristic HHG spectrum, consisting of a broad plateau of harmonic orders with comparable intensity followed by a sharp high-energy cutoff.
Electrons that do not recombine with their parent ion instead contribute to the partial ionization of the medium.

The macroscopic buildup of the harmonic field is governed by phase matching, which depends on both propagation-induced phase shifts and the phase of the microscopic atomic response. Accordingly, the wave-vector mismatch contains contributions from neutral-gas dispersion, plasma dispersion due to laser-induced free electrons, the intensity-dependent atomic dipole phase, and the geometric phase associated with focusing \cite{Constant1999,gaarde2008macroscopic,Heyl_2017,Cabasse12}.
The macroscopic HHG process also exhibits geometrical scale invariance, whereby the driving-laser pulse energy, focusing geometry, gas pressure, and interaction length can be scaled according to specific relations while preserving the harmonic conversion efficiency \cite{Heyl_2017}. Phase-matching and absorption considerations further reveal a characteristic hyperbolic relation between the pressure and medium length corresponding to optimal conversion efficiency \cite{weissenbilder2022optimize,Appi:23}.
A related dependence has recently been identified for the driving-pulse duration: at a given intensity, experiments and simulations show that the conversion efficiency is maximized at a particular pulse duration, and that this maximum scales approximately with the inverse square root of that duration
\cite{westerberg2025influence}.

While these scaling relations provide valuable guidance for identifying favorable operating conditions, maximizing the conversion efficiency requires the simultaneous optimization of several interdependent gas and laser parameters \cite{weissenbilder2022optimize}. 
Although low-dimensional parameter spaces can be explored using direct numerical scans, the computational cost increases rapidly with the number of optimization variables, making exhaustive exploration increasingly impractical.
Beyond identifying optimal conditions, reconstructing the conversion-efficiency landscape is valuable for understanding the dependence on different operating parameters and identifying extended regions of favorable conditions. 
This motivates approaches that can efficiently locate high-efficiency conditions while accurately reconstructing multidimensional HHG landscapes using a limited number of simulations.

Bayesian optimization (BO) provides an efficient framework for addressing such multidimensional problems. In recent years, it has increasingly been applied to computationally and experimentally expensive laser--matter interaction problems \cite{shalloo2020automation,jalas2021bayesian,BelmiroChu:21,YE2022106116,valenta2025bayesian,spears2025predicting,fan2025improved,arran2026bayesian,maslarova2026batch}, including waveform control and polarization gating for attosecond pulse generation \cite{BelmiroChu:21}. However, the potential of BO for efficiently optimizing and accurately reconstructing the multidimensional conversion-efficiency landscape in gas-based HHG remains largely unexplored.

BO constructs a probabilistic surrogate model of the quantity to be optimized from a limited number of evaluations and uses an acquisition function to iteratively select new points that are most informative or promising for locating the optimum \cite{BOreview,frazier2018tutorial}. 
By balancing exploration of uncertain regions with exploitation of promising ones, BO can identify high-performing conditions with substantially fewer evaluations than exhaustive parameter scans.
However, standard BO does not inherently account for known physical structure in the parameter space.
Commonly, Gaussian processes (GPs) \cite{rasmussen2006gaussian} are used as surrogate models, describing correlations between points based on their Euclidean distance in the chosen parameter space.
Consequently, the choice of coordinates influences how accurately the GP can represent the underlying physical landscape.

Motivated by known scaling relations associated with enhanced conversion efficiency, including the pressure--length hyperbola \cite{weissenbilder2022optimize,Appi:23} and the pulse-duration dependence \cite{westerberg2025influence}, we incorporate physical knowledge directly into the BO search space through nonlinear coordinate transformations.
The two-dimensional pressure--length landscape is sufficiently low-dimensional to permit a direct reference grid scan, providing a controlled setting for demonstrating how physics-informed coordinates improve surrogate reconstruction beyond simply locating the optimum. 
We then extend the approach to four dimensions by including the driving-laser intensity and pulse duration, where exhaustive scanning becomes computationally impractical and accurate reconstruction of the coupled parameter landscape from limited simulations becomes particularly valuable.
We demonstrate that even this relatively simple incorporation of physical knowledge substantially improves the global fidelity of the surrogate model.
Accurate surrogate reconstruction provides information beyond the location of the optimum, revealing the dependence of conversion efficiency on the operating parameters and enabling the identification of extended high-efficiency regions.

A more faithful surrogate provides additional information about the underlying parameter dependence and enables the identification of broad, high-efficiency operating regimes.
Beyond locating the global optimum, we further employ a threshold-targeting acquisition strategy to efficiently map regions in which the conversion efficiency exceeds a prescribed value, providing high-efficiency operating windows rather than isolated optimal points.

The remainder of this paper is organized as follows.
Section \ref{sec:methodology_hhg} outlines the numerical framework for modelling HHG, including the setup and input parameters used for the simulations.
Section \ref{sec:methodology_bo} describes the BO methodology, including the GP surrogate model, acquisition functions, and performance metrics.
Section \ref{sec:2D} presents the results of the two-dimensional optimization, evaluating different search-space parameterizations and demonstrating the identification of extended high-efficiency operating regions.
Section \ref{sec:4D} shows the results of moving to a four-dimensional search space, and
Section \ref{sec:conclusions} concludes.

\section{Methodology of modelling High-Harmonic Generation}\label{sec:methodology_hhg}
The HHG process is modelled numerically by coupling the microscopic single-atom response to the macroscopic propagation of both the driving IR field and the generated harmonic fields. 
In this work, we adopt the approach described in Refs.~\cite{lhuillier1992,LHuillier1996,harp_2026,Mattyspaper}.

\subsection{Numerical model}\label{subsec:model}
The macroscopic propagation of both the driving IR field and the generated harmonic fields is described using the paraxial Helmholtz equation under the slowly varying envelope approximation.
Spatial propagation is solved numerically in cylindrical coordinates, assuming cylindrical symmetry, using a finite-difference Crank--Nicolson scheme~\cite{crank1947practical}.
For the fundamental field, the propagation accounts for the linear atomic response, the contribution of laser-induced free electrons, and the Kerr nonlinearity \cite{kerr1875xl}. The field is initialized from the prescribed laser profile at each time slice and propagated through the medium.

The microscopic atomic response is obtained from precomputed TDSE tables \cite{carlstrom2022generala,carlstrom2022generalb, weissenbilder2022optimize}. These tables contain the intensity-dependent harmonic dipole amplitudes and phases, as well as the ionization rates. The local fundamental field is used to interpolate these quantities, while the ionization rates determine the evolution of the atomic and free-electron densities.
The harmonic dipole response is then used to construct the nonlinear polarization, which acts as the source term in the propagation equation for the generated harmonic fields. The propagation of the harmonic fields further includes dispersion and absorption through the atomic form factors obtained from the Henke tables \cite{henke1993x}.
All governing equations described above are implemented in the Python-based numerical framework HarP \cite{harp_2026}, publicly available through the HarmonicPropagation GitHub organization \cite{harmonicpropagation}, with its underlying physical model and numerical implementation detailed in Ref.~\cite{Mattyspaper}.

\subsection{Simulation setup}
Numerical simulations were performed using a common setup within the HarP code, varying only the physical parameters targeted during BO.
These variable parameters included the argon gas medium pressure $p$ and length $L$, the driving laser's peak intensity $I$, and its full width at half maximum (FWHM) pulse duration $\tau$.
All other gas properties and laser parameters were kept constant across all runs, while the numerical discretization followed the same strategy throughout.

The central wavelength of the driving laser pulse was $810~\mathrm{nm}$, with linear polarization and Gaussian temporal intensity and transverse spatial profiles.
The Rayleigh length (the distance from the beam waist to where its area doubles) was fixed at $z_R=18.6~\mathrm{mm}$, corresponding to a beam waist of $w_0=69.3~\mathrm{\upmu m}$.
The laser focus was positioned at the centre of the gas medium, $z_{\mathrm{focus}}=L/2$, while the temporal peak of the pulse was placed at the centre of the simulation window.
The argon gas was represented by a sharp-edged, spatially uniform profile with constant pressure $p$ from $z=0$ to $z=L$, and zero gas density outside this interval.
The atomic number density was determined via the ideal gas law, $\mathcal{N}=p/(k_{\mathrm{B}} T)$, where $k_{\mathrm{B}}$ is the Boltzmann constant and $T=293~\mathrm{K}$ is the room temperature.
The atomic polarizability was approximated by its static value $\alpha_0\simeq1.8\times10^{-40}~\mathrm{C\,m^2\,V^{-1}}$, since the dynamic polarizability is expected to remain close to $\alpha_0$ at the driving wavelength of $810\,\mathrm{nm}$ \cite{soldan2001static,Gaiser2018,Polyanskiy2024}.
The Kerr nonlinearity was included using the nonlinear refractive index of argon at $800~\mathrm{nm}$, $n_2=9.8\times10^{-24}~\mathrm{m^2\,W^{-1}}$ \cite{nibbering1997determination}.
Linear dispersion and absorption were calculated from the tabulated Henke atomic scattering factors \cite{henke1993x}, while ionization rates and harmonic dipole moments were taken from precomputed tables~\cite{carlstrom2022generala, carlstrom2022generalb,weissenbilder2022optimize}, as described in Subsection \ref{subsec:model}. In particular, tables utilized in this work were obtained from three-dimensional TDSE calculations within the single-active-electron approximation, using a laser field with a five-cycle ramp followed by 15 cycles of constant amplitude. The harmonic response was extracted by Fourier analysis of the dipole response over the final five optical cycles.

As the pressure--length hyperbolic relation for the conversion efficiency was previously investigated for the $23^\mathrm{rd}$ harmonic in Ref.~\cite{weissenbilder2022optimize}, the same quantity was selected as the objective function for the present optimization, defined as
\newcommand{\CE}{\ensuremath{\mathrm{CE}}}
\begin{equation}\label{eq:objective}
\CE=\frac{\mathcal{E}_{23}}{\mathcal{E}_{\mathrm{L}}},
\end{equation}
where $\mathcal{E}_{23}$ is the energy of the $23^\mathrm{rd}$ harmonic at the end of the medium and $\mathcal{E}_{\mathrm{L}}$ is the incident laser pulse energy. Assuming cylindrical symmetry, $\mathcal{E}_{23}$ was calculated by numerically integrating the absolute square of the propagated $23^\mathrm{rd}$ harmonic field over the temporal and transverse plane coordinates. The incident laser pulse energy was calculated assuming Gaussian spatial and temporal intensity profiles as $\mathcal{E}_{\mathrm{L}}=\pi^{3/2}Iw_0^2\tau/(4\sqrt{\ln2})$.

The computational domain extended to $L_r=3w_0$ in the radial direction and covered the entire gas-medium length $L$ in the longitudinal direction.
The length of the simulation box in the temporal direction was $L_{t}=3\tau$. 
Resolutions on radial and temporal grids were fixed to  $\Delta r=202.9~\mathrm{nm}$ and $\Delta t=128.9~\mathrm{as}$, respectively. 
Longitudinal sampling scaled with the medium length according to $\Delta z=L/256$. 
Numerical convergence tests validating the chosen grid parameters are available in the BO GitHub repository \cite{harp_bo_zenodo}.
The following section describes the BO methodology used to explore the conversion-efficiency landscape of the $23^\mathrm{rd}$ harmonic, based on the HHG simulation framework and parameters introduced above.
 
\section{Methodology of Bayesian optimization}\label{sec:methodology_bo}

Bayesian optimization combines a probabilistic surrogate model with an acquisition function to iteratively select new evaluation points \cite{BOreview,frazier2018tutorial}.
In the present work, the surrogate model approximates the CE from the completed HHG simulations and provides both a prediction and an associated uncertainty across the optimization space.
The acquisition function then uses this information to determine the next evaluation point.

Two optimization scenarios were considered in this work: a 2D optimization involving the gas pressure and medium length, and a 4D optimization in which the laser intensity and pulse duration were additionally varied.
In both cases, the parameter ranges were $3 \leq pz_R \leq 300~\mathrm{mbar\,cm}$ and $0.015 \leq L/z_R \leq 1.5$. In the 2D optimization, the laser intensity and pulse duration were fixed at $I=2.5\times10^{14}~\mathrm{W\,cm^{-2}}$ and $\tau=22~\mathrm{fs}$, respectively, whereas the 4D optimization additionally explored $10^{12} \leq I \leq 10^{15}~\mathrm{W\,cm^{-2}}$ and $11 \leq \tau \leq 88~\mathrm{fs}$. 
These ranges were selected to explore a broad range of gas and laser conditions while maintaining a consistent parameter domain for comparing the different BO strategies. 
The resulting optimization is therefore restricted to the prescribed domain, rather than intended to establish the absolute maximum CE over all physically accessible conditions.
Since both studies employ the same BO methodology, the common BO workflow, surrogate model, and acquisition functions are presented below.

\subsection{BO workflow}
The complete BO framework was implemented in Python using the BoTorch \cite{balandat2020botorch} and GPyTorch \cite{gardner2018gpytorch} libraries. The resulting HarP-BO framework \cite{harp_bo_zenodo}, also available through the HarmonicPropagation GitHub organization \cite{harmonicpropagation}, interfaces the optimizer with the HarP simulation code \cite{harp_2026}.
\begin{figure}[b]
    \centering
    \vspace{5mm}
    \input{bo_scheme}
    \caption{\textbf{BO scheme.} Schematic illustration of the Bayesian optimization workflow. Following the initial CE evaluation, the GP surrogate is updated and the acquisition function is optimized to select the next set of simulation parameters. The HHG simulation is then performed, and the resulting CE is used to update the surrogate model. This procedure is repeated for $N_{\mathrm{BO}}$ iterations.}
    \label{fig:BOscheme}
\end{figure}
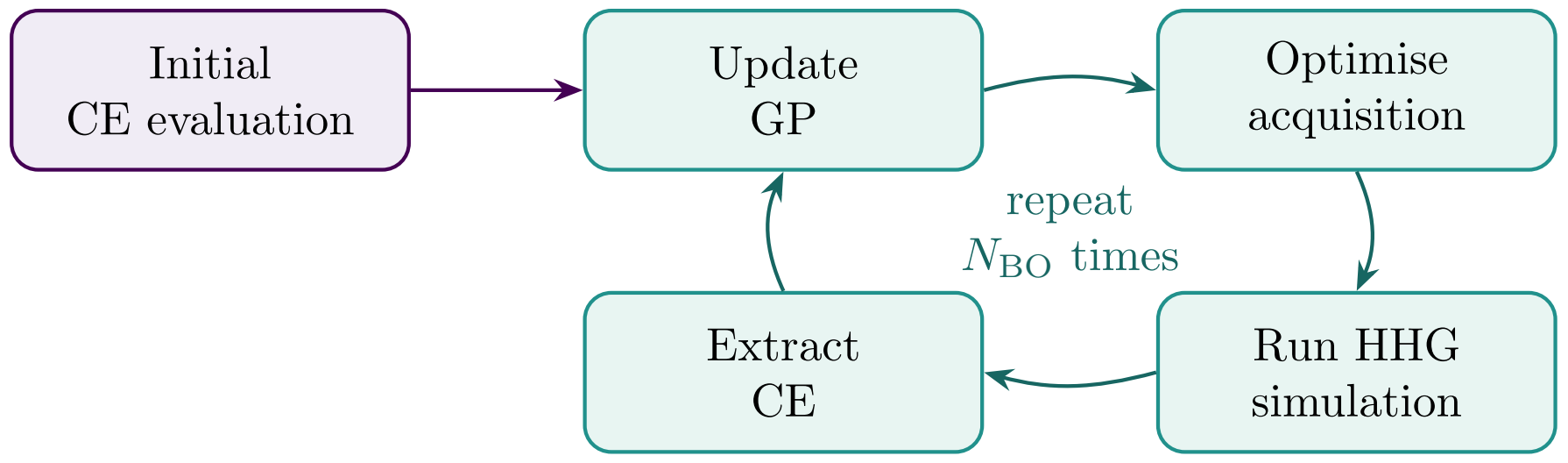

The overall optimization procedure employed in this work is illustrated in Figure~\ref{fig:BOscheme}.
The optimization begins with an initial HHG simulation performed at an initial point generated from a randomly seeded uniform draw. 
After the corresponding CE has been evaluated, a surrogate model is constructed.
In the present work, a GP \cite{rasmussen2006gaussian} is employed as the surrogate model.
The acquisition function is then evaluated using the GP posterior to identify the next point in the optimization space.
A new HHG simulation is subsequently performed at the selected point, and the resulting CE is extracted from the simulation output and incorporated into the training dataset.
The GP is then updated using the new simulation, after which the acquisition function is evaluated again to determine the next simulation point.
This optimization cycle is repeated for $N_\mathrm{BO}$ BO iterations, corresponding to $N_\mathrm{BO}$ HHG simulations selected by the BO algorithm.
Including the first simulation used to initialize the optimization, each optimization consequently consisted of $1+N_\mathrm{BO}$ HHG simulations in total.

\subsection{GP surrogate model and covariance kernel}\label{sec3:kernels}
Before GP fitting, the CE was normalised by a reference value of $10^{-6}$. The resulting objective was either used directly or subjected to an additional square-root transformation, depending on the optimization configuration.
The objective values were then standardized to have zero mean and unit variance before fitting the GP.
The objective modeled by the GP is denoted by $f(\mathbf{x})$.
The GP provides a posterior mean $\mu_f^i(\mathbf{x})$ and a posterior standard deviation $\sigma_f^i(\mathbf{x})$ at every point $\mathbf{x}=(x_1,\ldots,x_D)$ in the $D$-dimensional optimization space.

For GP fitting, each optimization coordinate, including any nonlinear transformations, was linearly rescaled to the interval $[0,1]$ using its corresponding minimum and maximum values. The coordinates $\mathbf{x}$ in the following GP expressions refer to this normalized optimization space.
The iteration index $i=0$ denotes the GP fitted to the initial CE evaluation, while $i\geq 1$ denotes subsequent BO iterations.
To assess the sensitivity of the BO results to random initialization, all 2D and 4D optimizations presented in the main text were repeated using five independent random seeds.
The figures in the main text illustrate individual optimization runs, while the mean and standard deviation of the optimization convergence and GP reconstruction accuracy across the five seeds are provided in the HarP-BO repository \cite{harp_bo_zenodo}.
These additional results support the overall trends observed in the main text, indicating that the improvements obtained with physics-informed parameterizations are not specific to a particular initialization.
The corresponding datasets are also available in the repository for reproducibility and further analysis.

The covariance $K$ between two points $\mathbf{x}=(x_1,\ldots,x_D)$ and $\mathbf{x}'=(x'_1,\ldots,x'_D)$ in the optimization space was described using a Matérn $5/2$ kernel \cite{matern1960spatial, rasmussen2006gaussian}:
\begin{equation}
K_{\theta}(\mathbf{x},\mathbf{x}')=s^2\left(1+\sqrt{5}R_{\boldsymbol{\ell}}+\frac{5}{3}R_{\boldsymbol{\ell}}^2\right)\exp\left(-\sqrt{5}R_{\boldsymbol{\ell}}\right),
\label{eq:matern52}
\end{equation}
 where $R_{\boldsymbol{\ell}}$ is the scaled Euclidean distance between the two points defined as
\begin{equation}
R_{\boldsymbol{\ell}}=\sqrt{\sum_{j=1}^{D}\frac{\left(x_j-x_j'\right)^2}{\ell_j^2}}.
\label{eq:matern_distance}
\end{equation}

\noindent The kernel hyperparameters are denoted by $\theta=(s^2,\boldsymbol{\ell})$, where $s^2$ is the output variance, which controls the overall magnitude of the variations represented by the GP, and $\boldsymbol{\ell}=(\ell_1,\ldots,\ell_D)$ is the vector of characteristic length scales, with $\ell_j$ corresponding to the $j^{\mathrm{th}}$ coordinate of the optimization space. For coincident input points, $\mathbf{x}=\mathbf{x}'$, the scaled distance vanishes and the kernel reduces to $K_\theta(\mathbf{x},\mathbf{x})=s^2$.

The Matérn $5/2$ kernel assumes that the objective function is twice mean-square differentiable, providing a balance between smooth variations and more localized features of the CE landscape.
To assess the influence of the kernel choice, we also compared the Matérn $1/2$, Matérn $3/2$, and radial basis function (RBF) kernels in the 2D optimization using the physics-informed $(\log(pL),\log(p/L))$ parameterization.
These kernels differ in their assumptions about smoothness: the Matérn $1/2$ describes non-differentiable functions, the Matérn $3/2$ assumes once mean-square differentiable functions, and the RBF kernel assumes infinitely differentiable functions.
The Matérn $3/2$, Matérn $5/2$, and RBF kernels showed similar optimization convergence, while the Matérn $5/2$ yielded slightly better final surrogate reconstruction accuracy.
The Matérn $1/2$ run was discontinued after 10 iterations due to GP fitting failure.
Based on these results, the Matérn $5/2$ kernel was selected for the main optimization studies.
The complete kernel comparison, including convergence curves and reconstruction metrics, is available in the HarP-BO repository \cite{harp_bo_zenodo}.

Automatic relevance determination (ARD) \cite{neal1996bayesian, rasmussen2006gaussian} was used to assign an independent characteristic length scale $\ell_j$ to each optimization coordinate, allowing the GP to capture differences in the sensitivity of the objective function to individual parameters.
The kernel hyperparameters $\theta=(s^2,\boldsymbol{\ell})$ were re-estimated at each BO iteration by maximizing the marginal likelihood, which measures how well the GP model explains the observed simulation data.
Since HarP is deterministic, the GP observation-noise variance was fixed at $10^{-5}$, much smaller than the unit variance of the standardized objective values, rather than fitted.
The length scales and output variance were initialized to approximately $0.693$ using the default GPyTorch initialization.
To improve fitting stability, the fitted GP parameters from the preceding iteration were also considered as an alternative starting point, retaining the fit with the higher marginal likelihood.
The resulting GP posterior mean $\mu_f^i(\mathbf{x})$ and standard deviation $\sigma_f^i(\mathbf{x})$ were then used to construct the acquisition function for selecting the next simulation point.

\subsection{Acquisition function}
The role of the acquisition function is to balance exploration of uncertain regions of the parameter space with exploitation of regions predicted to satisfy the optimization objective.
In conventional BO, the acquisition function is maximized directly in the optimization space subject to the prescribed variable bounds.
However, when nonlinear coordinate transformations are employed, these bounds no longer coincide with the admissible region of the original HHG parameter space.
Consequently, direct optimization of the acquisition function may propose parameter combinations that lie outside the intended physical search region.

To ensure that all proposed simulation points remain within the intended physical search region, candidate points were randomly generated in the normalized optimization space and transformed back to the original physical coordinates.
Candidates violating the prescribed bounds of any optimized parameter ($p$ and $L$ in the 2D case and $p$, $L$, $I$ and $\tau$ in the 4D case) were rejected, and sampling continued until a pool of 512 physically valid candidates was obtained.
The acquisition function was then evaluated at these candidates, and the point with the highest acquisition value was selected for the next HHG simulation.

In this work, two complementary acquisition strategies were investigated.
Log Expected Improvement (LogEI) \cite{jones1998efficient, ament2023unexpected} was used to maximize the CE, whereas the Straddle acquisition function was employed to identify regions where the CE crossed a prescribed threshold.
The expected improvement (EI) acquisition function \cite{jones1998efficient} evaluates the expected gain of the CE over the best value observed so far, while accounting for predictive uncertainty.
For a GP surrogate model, standard EI is given by
\begin{equation}
a_{\mathrm{EI}}^i(\mathbf{x})=
(\mu_f^i(\mathbf{x})-f_{\mathrm{best}}^i)\Phi[\gamma^i(\mathbf{x})]
+\sigma_f^i(\mathbf{x})\phi[\gamma^i(\mathbf{x})],
\end{equation}
with
\begin{equation}
\gamma^i(\mathbf{x})=
\frac{\mu_f^i(\mathbf{x})-f_{\mathrm{best}}^i}
{\sigma_f^i(\mathbf{x})}.
\end{equation}
Here, $\gamma^i(\mathbf{x})$ measures how far the predicted mean lies above or below the best objective value observed so far, $f_{\mathrm{best}}^i$, scaled by the GP posterior standard deviation at the $i^{\mathrm{th}}$ iteration.
The functions $\Phi$ and $\phi$ represent the cumulative distribution and probability density functions of the standard normal distribution, respectively, defined as
$\phi(\xi)=(1/\sqrt{2\pi})\exp(-\xi^2/2)$
and
$\Phi(\xi)=\int_{-\infty}^{\xi}\phi(u)\,\mathrm{d}u$.
To prevent numerical underflow in regions of high certainty or poor performance, the numerically stable logarithmic variant, LogEI [$a_{\mathrm{LogEI}}^i(\mathbf{x}) = \log a_{\mathrm{EI}}^i(\mathbf{x})$], was employed \cite{ament2023unexpected} via the BoTorch library.

Unlike LogEI, which seeks to maximize the CE, the Straddle acquisition function aims to identify the boundary separating regions above and below a prescribed CE threshold.
This allows BO to efficiently map the high-efficiency region rather than locate a single optimum.
The Straddle acquisition function is defined as
\begin{equation}
a_{\mathrm{Str}}^i(\mathbf{x})=\beta \sigma_f^i(\mathbf{x})-\left|\mu_f^i(\mathbf{x})-f_{\mathrm{target}}\right|,
\label{eq:straddle}
\end{equation}
where $f_{\mathrm{target}}$ is the prescribed CE threshold expressed in the transformed objective space, and $\beta$ controls the balance between exploration and exploitation.
The first term promotes sampling in regions of high predictive uncertainty, whereas the second term favours points whose predicted objective value is close to the prescribed threshold.
In this work, $\beta=1.96$ was used, corresponding to an approximate 95\% confidence interval under the Gaussian posterior assumption.
Since the objective transformation preserves the ordering of CE values, the threshold in the transformed objective space corresponds to the same physical CE threshold.
As a result, the algorithm preferentially samples points near the estimated boundary of the high-efficiency region, allowing the region satisfying
\begin{equation}\label{eq:target}
\CE(\mathbf{x}) \geq \CE_{\mathrm{target}}
\end{equation}
to be identified efficiently.
The Straddle acquisition function was implemented specifically for the present work within the developed optimization framework \cite{harp_bo_zenodo}.

\subsection{Performance of BO}

The performance of BO was evaluated using three complementary measures: the convergence of the optimization, the normalized root-mean-square error (NRMSE), and the structural similarity index (SSIM) \cite{wang2004image}.
The convergence was assessed from the evolution of the best-so-far CE with BO iteration, whereas NRMSE was used to quantify the accuracy of the GP surrogate model relative to reference HHG simulations.
For the 2D optimization, SSIM was additionally employed to assess the structural similarity between the predicted and reference CE response surfaces, including features such as peaks, ridges, and valleys.
This measure was restricted to the 2D study, where the regular reference grid allows the overall landscape structure to be compared directly, whereas the 4D reference consists of scattered Sobol-sampled points.

For the 2D optimization, the reference CE response surface was obtained from a direct $100\times100$ grid scan uniformly covering the physical parameter domain.
After each BO iteration, the GP posterior mean was evaluated at the same grid points and compared with the reference surface.
For the 4D optimization, where a dense grid scan becomes computationally expensive, an independent reference dataset was generated using a scrambled Sobol sequence consisting of 1024 points \cite{sobol1967distribution}.
Sobol sequences are quasi-random sequences designed to distribute points more uniformly throughout a multidimensional space than conventional random sampling.
This reduces clustering and large gaps between sampled regions, allowing the GP predictions to be tested across the four-dimensional parameter domain.
The same reference points were used to evaluate all GP parameterizations.

The NRMSE is defined as
\begin{equation}
\mathrm{NRMSE}(G,P^i)=
\frac{
\sqrt{\frac{1}{M}\sum_{m=1}^{M}\left(G_m-P_m^{i}\right)^2}
}{D},
\label{eq:nrmse}
\end{equation}
where $G_m$ and $P_m^i$ are the reference CE obtained from the HarP simulation and the GP-predicted CE at the same parameter combination, respectively.
The index $m$ labels the reference points ($M=10\,000$ for the 2D grid scan and $M=1024$ for the 4D Sobol dataset), while $i$ denotes the BO iteration.
The normalization factor
\begin{equation}
D=\max(G)-\min(G)
\end{equation}
is the dynamic range of the reference CE values.
For the 2D optimization, NRMSE was evaluated on the direct grid scan ($\mathrm{NRMSE}_{\mathrm{grid}}$), whereas for the 4D optimization, it was evaluated on the Sobol-sampled reference dataset ($\mathrm{NRMSE}_{\mathrm{Sobol}}$).

The SSIM, evaluated only for the 2D optimization on the direct grid scan ($\mathrm{SSIM}_{\mathrm{grid}}$), is defined in terms of the mean, variance, and covariance:
\begin{equation}
\mathrm{SSIM}(G,P^{i})=
\frac{
\left[2\,\operatorname{Mean}(G)\operatorname{Mean}(P^{i})+C_1\right]
\left[2\,\operatorname{Cov}(G,P^{i})+C_2\right]
}{
\left[\operatorname{Mean}(G)^2+\operatorname{Mean}(P^{i})^2+C_1\right]
\left[\operatorname{Var}(G)+\operatorname{Var}(P^{i})+C_2\right]
}.
\label{eq:ssim}
\end{equation}
SSIM was originally developed for image analysis and is typically evaluated locally using sliding windows \cite{wang2004image}.
Here, it was computed globally across the entire 2D parameter space to assess the overall structural agreement between the reference and predicted CE landscapes without dividing them into smaller regions.
The constants $C_1=(c_1D)^2$ and $C_2=(c_2D)^2$ prevent numerical instabilities for small denominators, with $c_1=0.01$ and $c_2=0.03$ chosen following Ref.~\cite{wang2004image} to provide stabilization while limiting their influence on the similarity measure.

A lower NRMSE indicates closer numerical agreement between the predicted and reference CE values, with a value of zero indicating identical predictions.
A higher SSIM indicates better structural agreement between the predicted and reference surfaces, with a value of 1 corresponding to perfect structural agreement.
Together, these metrics distinguish between the ability of BO to locate the optimum and the ability of the GP surrogate to accurately reproduce the CE landscape.

\section{Results of 2D optimization}\label{sec:2D}

In this section, we investigate the two-dimensional pressure--length optimization using two complementary acquisition strategies. We first compare different search-space parameterizations using LogEI, focusing on optimization convergence and GP surrogate reconstruction accuracy. We then employ the Straddle acquisition function to identify extended high-efficiency operating regions beyond the isolated optimum.

\subsection{Comparison of search-space parameterizations}

The optimal CE in the pressure--length space follows an approximately hyperbolic ridge, as predicted by theory \cite{Constant1999, weissenbilder2022optimize} and confirmed by previous numerical simulations \cite{weissenbilder2022optimize}.
To facilitate its reconstruction by a GP with a Matérn $5/2$ kernel, we investigate nonlinear coordinate transformations that better align the search space with this physical structure.

First, we introduce scaled coordinates based on the geometrical scaling of HHG: $pz_R$ for pressure and $L/z_R$ for length \cite{Heyl_2017}.
For simplicity, we use the notation $p$ and $L$ for these scaled quantities when describing the optimization parameterizations, while the figures retain the explicit scaling with $z_R$.
Second, we investigate nonlinear transformations of these coordinates to map the hyperbolic ridge into a geometry more easily represented by the stationary Matérn $5/2$ GP kernel, whose correlations depend on the scaled distance between points in the optimization space. The five studied parameterizations are:
\begin{equation}
\begin{aligned}
&(p,L),    && (\log p,\log L), \\
&(pL,p/L), && (\sqrt{pL},\sqrt{p/L}), && (\log(pL),\log(p/L)),
\end{aligned}
\end{equation}
where $\log$ denotes the base-10 logarithm. The first two parameterizations, $(p,L)$ and $(\log p,\log L)$, serve as reference cases for comparison with the physics-informed parameterizations. The former uses the original physical coordinates, while the latter introduces logarithmic scaling to account for the large ranges of pressure and medium length.
The remaining parameterizations exploit the approximately hyperbolic ridge, characterized by $pL\approx\mathrm{const}$, through combinations of the product and ratio $(pL,p/L)$. 
The product $pL$ identifies a particular hyperbola, while the ratio $p/L$ distinguishes different pressure--length combinations along it.
Although either $p$ or $L$ could serve as the second coordinate, the ratio $p/L$ describes this variation symmetrically, without favoring either parameter.
We consider three representations of these product--ratio coordinates: $(pL,p/L)$, $(\sqrt{pL},\sqrt{p/L})$, and $(\log (pL), \log (p/L))$.
All three straighten the hyperbolic ridge but differ in the scaling of distances between points in the GP kernel. 
The square-root and logarithmic transformations progressively compress the coordinate ranges, altering how distances between points are represented by the GP kernel.
We therefore compare these transformations to investigate how the choice of coordinates affects the GP reconstruction.

\begin{figure}[t!]
    \centering
    \includegraphics[width=1\linewidth]{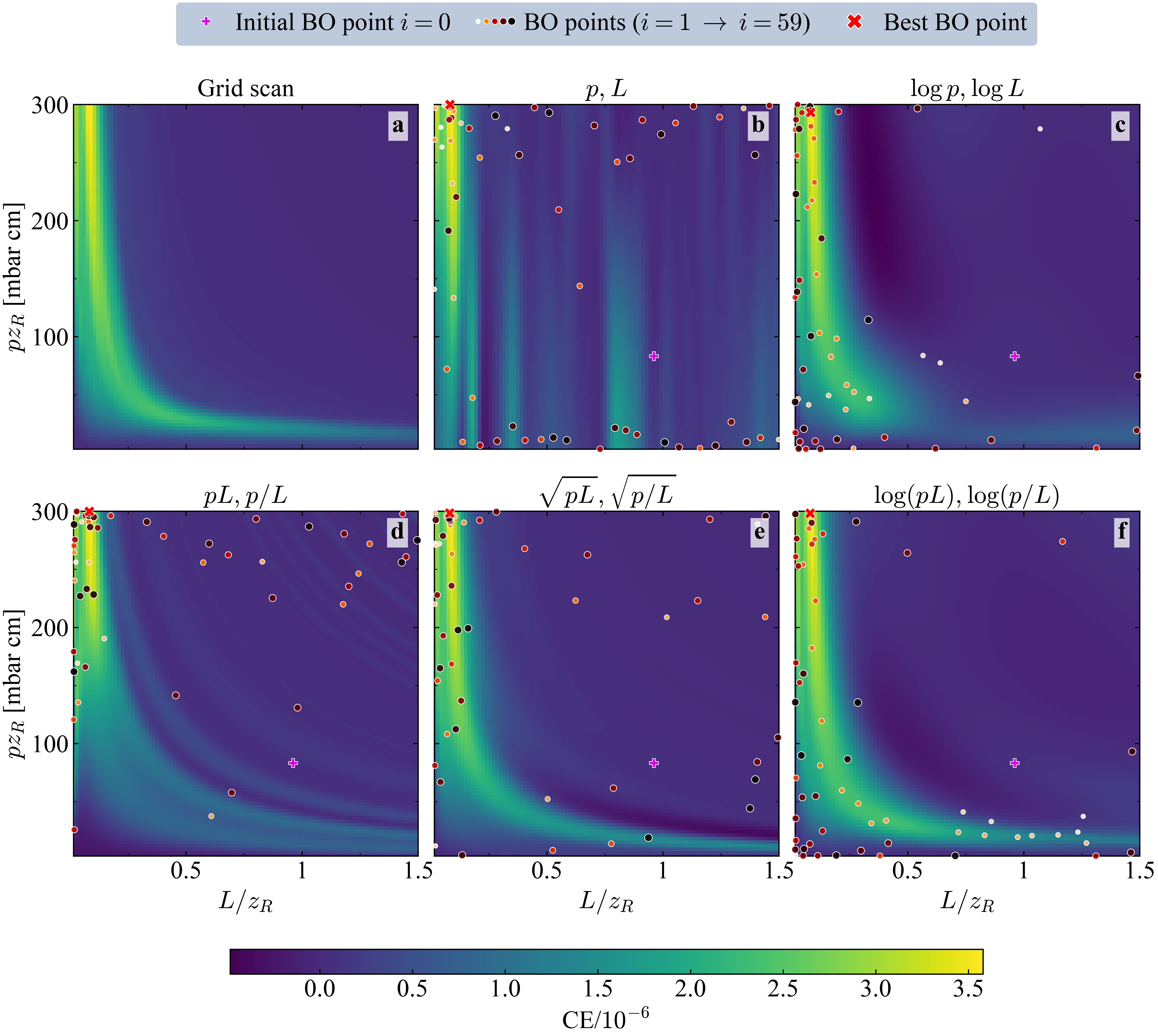}
\caption{\textbf{GP prediction of CE for different optimization parameterizations depicted in the scaled pressure--length coordinates $(pz_R,L/z_R)$.}
\textbf{a)} Direct grid scan of CE as a function of the pressure $p$ and gas-cell length $L$.
\textbf{b--f)} GP model predictions after 59 BO iterations for different optimization parameterizations:
\textbf{b)} $(p,L)$,
\textbf{c)} $(\log p, \log L)$,
\textbf{d)} $(pL,p/L)$,
\textbf{e)} $(\sqrt{pL},\sqrt{p/L})$, and
\textbf{f)} $(\log(pL),\log(p/L))$.
The magenta plus depicts the first shared initial evaluation, while circles indicate subsequent BO evaluations through iteration $i=59$. The circle size increases and the circle color changes from light to dark with increasing BO iteration number (1 $\rightarrow$ 59).
In b)--f), the red cross marks the best-performing point found by BO.
All panels share the same color scale and have a grid resolution of $100\times100$. Small negative values in the GP-predicted CE arise from the unconstrained GP surrogate and do not represent physically negative conversion efficiencies.}
    \label{fig:2D_ei_fig1}
    \end{figure}

\begin{figure}[h]
    \centering
        \vspace{2.5mm}
    \includegraphics[width=0.965\linewidth]{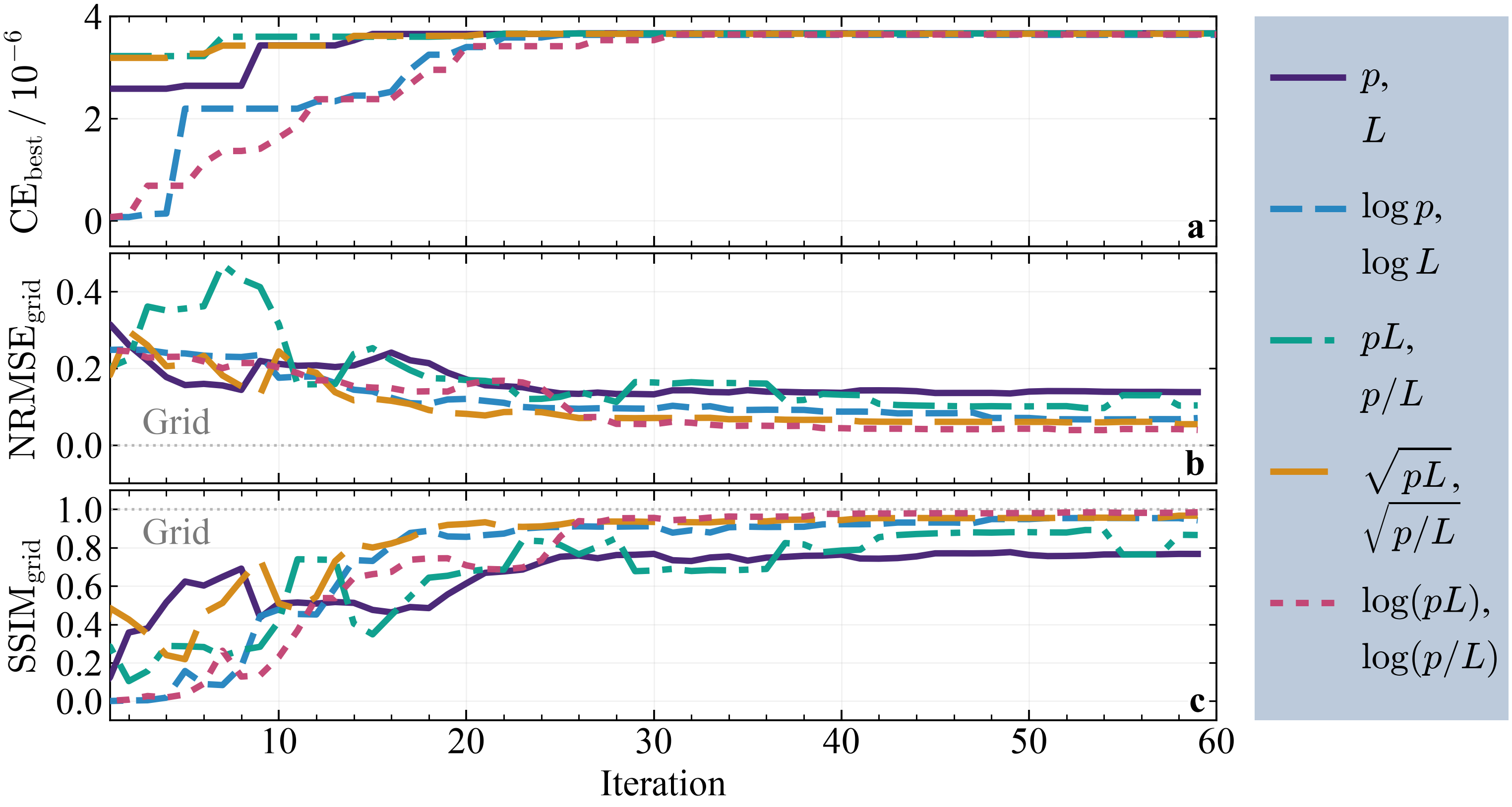}
    \caption{\textbf{Convergence and GP prediction accuracy for different 2D optimization parameterizations.} 
\textbf{a)} Evolution of the best-so-far CE as a function of BO iteration.
 \textbf{b)} NRMSE between the GP prediction and the direct grid scan ($\mathrm{NRMSE}_{\mathrm{grid}}$) as a function of BO iteration. The dotted gray horizontal line indicates the value of $0$, corresponding to perfect agreement with the direct grid scan.
\textbf{c)} SSIM between the GP prediction grid and the direct grid scan ($\mathrm{SSIM}_{\mathrm{grid}}$) as a function of BO iteration. The dotted gray horizontal line indicates the value of $1$, corresponding to perfect agreement with the direct grid scan. Different colors and marker styles distinguish the optimization parameterizations. }
    \label{fig:2D_ei_fig2}
    \end{figure}

For each parameterization, BO was performed for 60 iterations following a shared initial evaluation at $i=0$.
Since the GP surrogate is fitted before each subsequent BO evaluation, the GP predictions and reconstruction metrics correspond to observations through iteration $i=59$, whereas the best-so-far CE includes the evaluation at $i=60$.
Figures~\ref{fig:2D_ei_fig1} and \ref{fig:2D_ei_fig2} compare the GP surrogate reconstruction and optimization convergence for the five parameterizations.

%
Figure~\ref{fig:2D_ei_fig1}a shows the reference $100\times100$ direct grid scan, revealing the hyperbolic ridge of high CE.
In the standard linear $(p,L)$ space (Figure~\ref{fig:2D_ei_fig1}b), the GP fails to reproduce this hyperbolic structure, yielding a fragmented surrogate reconstruction with pronounced vertical bands.
Applying a logarithmic transformation, $(\log p,\log L)$ in Figure~\ref{fig:2D_ei_fig1}c, substantially improves the surrogate reconstruction, although small deviations from the reference landscape remain, particularly along the low-pressure branch of the hyperbolic ridge.
The product--ratio parameterization $(pL,p/L)$ (Figure~\ref{fig:2D_ei_fig1}d) captures the underlying hyperbolic geometry but produces multiple artificial ridge-like structures, deviating from the single dominant high-efficiency ridge observed in the reference grid scan. 

In contrast, combining the product--ratio representation with square-root or logarithmic transformations, as shown in Figures~\ref{fig:2D_ei_fig1}e and f, allows the GP to reproduce the high-efficiency ridge more accurately.
The comparison reveals that accurate reconstruction benefits from both the scaling and the physical structure of the optimization coordinates. 
Logarithmic scaling alone substantially improves the GP prediction, while the product--ratio coordinates capture the characteristic hyperbolic ridge but introduce artificial structures when used without further transformation. 
Combining the two provides the most accurate reconstruction, suggesting that both appropriate scaling and alignment with the underlying physical geometry are important for the GP surrogate.

The choice of parameterization influences surrogate reconstruction in two ways: first, by changing how the 512 candidate points are distributed in the original pressure--length space and, second, by changing how the GP models correlations between points, which in turn affects the acquisition function. The observed differences in reconstruction accuracy therefore reflect the combined influence of these effects.

Figure~\ref{fig:2D_ei_fig2} quantifies the optimization convergence and GP prediction accuracy across the different coordinate parameterizations.
All parameterizations reach similar maximum CE values, in the range of $3.6$--$3.7\times10^{-6}$.
However, in the runs shown, their convergence rates differ, with the $(pL,p/L)$ and $(\sqrt{pL},\sqrt{p/L})$ parameterizations converging most rapidly, followed closely by the standard linear $(p,L)$ coordinates.
The $(\log(pL),\log(p/L))$ and $(\log p,\log L)$ parameterizations approach similar maximum CE values more gradually.

While all parameterizations achieve similar maximum CE values within 60 BO iterations, Figures~\ref{fig:2D_ei_fig2}b and c reveal substantial differences in the global predictive accuracy of the GP surrogate models.
The $(\log(pL),\log(p/L))$ parameterization achieves the highest final surrogate accuracy, with $\mathrm{NRMSE}_{\mathrm{grid}}\approx0.04$ and $\mathrm{SSIM}_{\mathrm{grid}}\approx0.98$.
The $(\sqrt{pL},\sqrt{p/L})$ parameterization follows closely, with corresponding values of $0.05$ and $0.97$, respectively.
The similarly high accuracy obtained with the square-root product--ratio coordinates indicates that the improvement is not exclusive to logarithmic scaling, although the logarithmic representation provides the best overall reconstruction among those tested.
In contrast, the $(pL,p/L)$ and standard linear $(p,L)$ parameterizations exhibit lower global predictive accuracy, despite achieving comparable maximum CE values.

Beyond achieving comparable maximum conversion efficiencies, the physics-informed parameterizations provide substantially more accurate reconstructions of the CE landscape.
For example, the standard linear $(p,L)$ and $(\log(pL),\log(p/L))$ parameterizations achieve similar maximum CE values of approximately $3.67\times10^{-6}$ and $3.64\times10^{-6}$, respectively, but differ substantially in surrogate accuracy.
The former yields $\mathrm{NRMSE}_{\mathrm{grid}}\approx0.14$ and $\mathrm{SSIM}_{\mathrm{grid}}\approx0.77$, compared with $0.04$ and $0.98$ for the latter.

These results demonstrate that incorporating physical knowledge into the optimization coordinates substantially improves the reconstruction of the CE landscape using a limited number of simulations.
The resulting GP surrogate captures the characteristic hyperbolic structure while identifying maximum conversion efficiencies comparable to those obtained with conventional parameterizations.
This improved reconstruction is particularly valuable for exploring favorable operating regions and extending the approach to higher-dimensional parameter spaces where exhaustive scanning becomes computationally impractical.

\subsection{Identification of robust operating regimes}
\label{sec:robust_regimes}

While maximizing CE identifies the most favorable operating point, experimental applications also benefit from extended parameter regions that maintain sufficiently high efficiency, providing greater tolerance to variations in experimental conditions. To identify such operating windows, we employ the Straddle acquisition function described in Equation~\eqref{eq:straddle}, within the physics-informed $(\log(pL),\log(p/L))$ coordinate space, which achieved the highest surrogate reconstruction accuracy in the preceding LogEI comparison, as quantified by NRMSE and SSIM.

\begin{figure}[h!]
    \centering
    \includegraphics[width=1\linewidth]{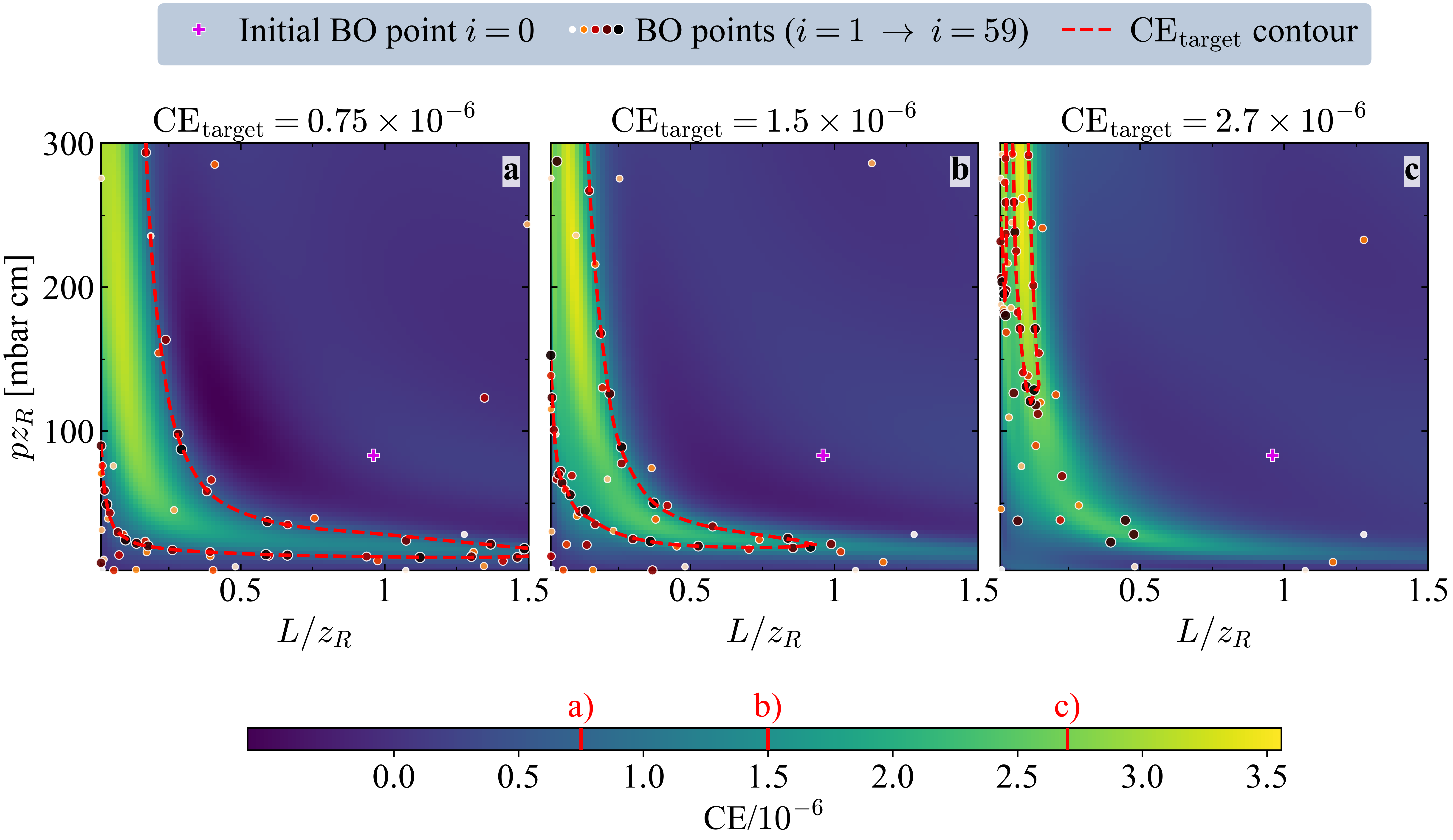}
\caption{\textbf{GP prediction of CE for different values of $\CE_{\mathrm{target}}$:}
\textbf{a)} $\CE_{\mathrm{target}}=0.75\times 10^{-6}$,
\textbf{b)} $\CE_{\mathrm{target}}=1.5\times 10^{-6}$,
\textbf{c)} $\CE_{\mathrm{target}}=2.7\times 10^{-6}$.
The magenta plus depicts the first shared initial evaluation, while circles indicate subsequent BO evaluations.
The circle size increases and the circle color changes from light to dark with increasing BO iteration number (1 $\rightarrow$ 60).
The red dashed line marks the region where the condition from Equation~\eqref{eq:target} is satisfied.
All panels share the same color scale and have a grid resolution of $100\times100$, corresponding to the resolution of Figure \ref{fig:2D_ei_fig1}. Small negative values in the GP-predicted CE arise from the unconstrained GP surrogate and do not represent physically negative conversion efficiencies.}
    \label{fig:2D_straddle_fig1}
\end{figure}

The target values used for the Straddle study were chosen with reference to the CE range obtained in the preceding LogEI comparison, where the highest efficiency across the tested parameterizations was approximately $3.7\times10^{-6}$. We selected three representative thresholds within this range to probe low, intermediate, and high CE levels.
Figure~\ref{fig:2D_straddle_fig1} shows the resulting GP surrogate predictions and sequential evaluation distributions after $N_{\mathrm{BO}}=60$ iterations for a low target, $\CE_{\mathrm{target}}=0.75\times10^{-6}$ (a), an intermediate target, $1.5\times10^{-6}$ (b), and a high target, $2.7\times10^{-6}$ (c).

Guided by the Straddle strategy, the BO algorithm systematically concentrates candidate evaluations along the boundary enclosing the target efficiency region (marked by the red dashed contours).
For the lower target values of $\CE_{\mathrm{target}}=0.75\times10^{-6}$ and $1.5\times10^{-6}$ (Figures~\ref{fig:2D_straddle_fig1}a and b), the high-efficiency region extends along both the long-medium/low-pressure and short-medium/high-pressure branches of the hyperbola.
As the threshold increases from $0.75\times10^{-6}$ to $1.5\times10^{-6}$, this region becomes narrower, particularly along the long-medium/low-pressure branch.
At the highest threshold of $2.7\times10^{-6}$ (Figure~\ref{fig:2D_straddle_fig1}c), the region narrows further to a small zone along the hyperbolic ridge, with sequential evaluations clustering densely around the short-medium/high-pressure parameter regime.
The additional narrow structure visible in the target contours may indicate a competing hyperbola branch alongside the dominant hyperbolic ridge.
This suggests that physics-informed parameterizations may also be beneficial in configurations exhibiting multiple hyperbolic branches, provided that these retain approximately hyperbolic pressure--length dependencies.
Assessing their performance in such regimes remains an interesting direction for future investigation.

\begin{figure}[h!]
    \centering
    \includegraphics[width=1\linewidth]{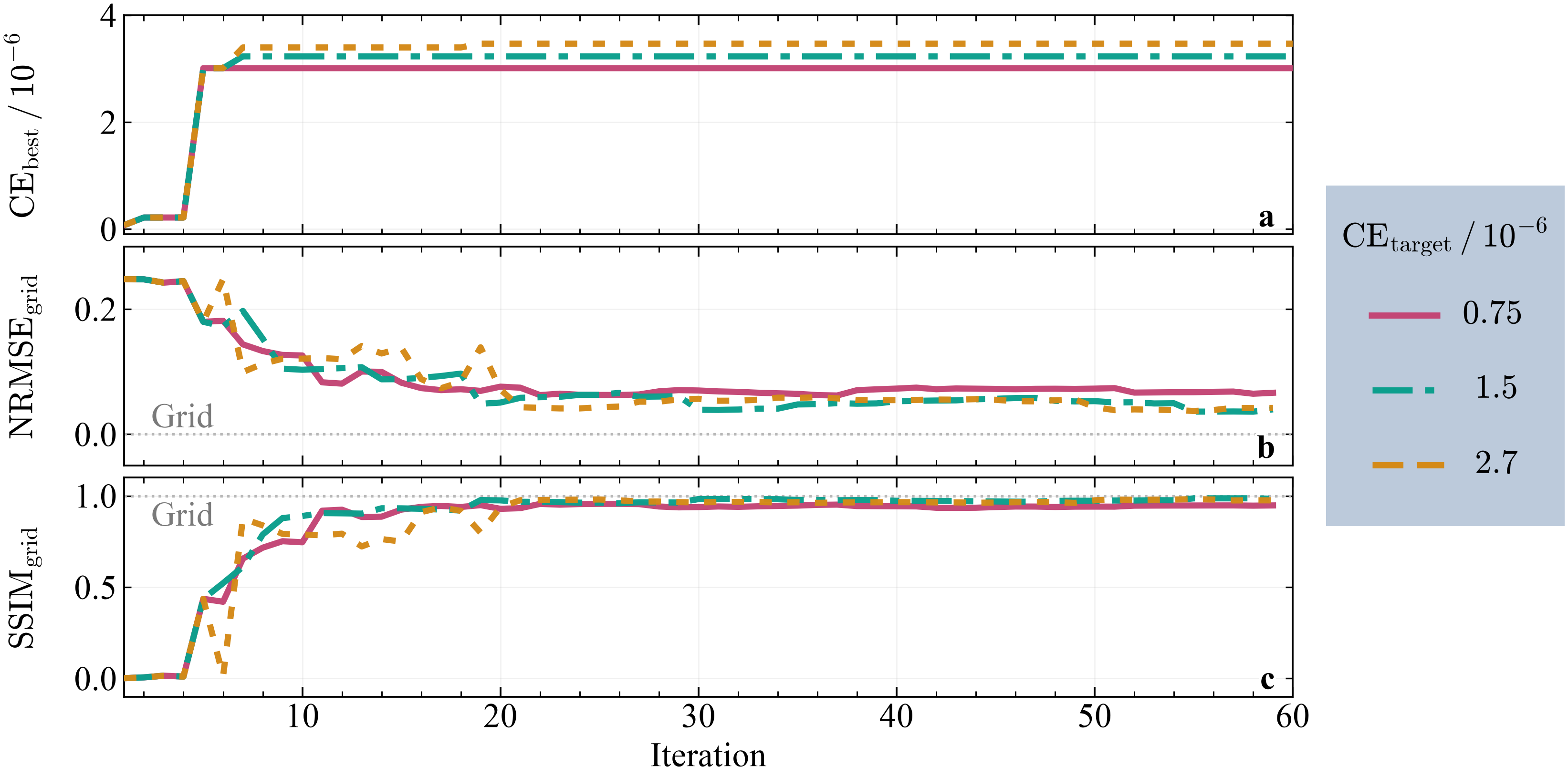}
    \caption{\textbf{Convergence and GP prediction accuracy for different values of $\CE_{\mathrm{target}}$.}
\textbf{a)} Evolution of the best-so-far CE as a function of BO iteration.
 \textbf{b)} NRMSE between the GP prediction and the direct grid scan as a function of BO iteration. The dotted gray horizontal line indicates the value of $0$, corresponding to perfect agreement with the direct grid scan. Note that the NRMSE axis is shown over a narrower range compared to Figure~\ref{fig:2D_ei_fig2} to better resolve the differences between the Straddle runs.
\textbf{c)} SSIM between the GP prediction grid and the direct grid scan as a function of BO iteration. The dotted gray horizontal line indicates the value of $1$, corresponding to perfect agreement with the direct grid scan. Different colors and marker styles distinguish different values $\CE_{\mathrm{target}}$. }
    \label{fig:2D_straddle_fig2}
\end{figure}

Figure~\ref{fig:2D_straddle_fig2} shows the optimization convergence and GP predictive accuracy for the three target thresholds. As shown in Figure~\ref{fig:2D_straddle_fig2}a, all three runs reach their highest evaluated CE within the first 20 iterations, despite the primary objective of Straddle being the identification of target-efficiency boundaries rather than direct maximization of CE. For the highest threshold ($\CE_{\mathrm{target}}=2.7\times10^{-6}$), the optimization concentrates around the core of the hyperbola ridge, reaching a CE of approximately $3.47\times10^{-6}$ after 19 iterations, close to the maximum obtained from the direct grid scan ($3.56\times10^{-6}$).

Moreover, across all values of $\CE_{\mathrm{target}}$, the Straddle acquisition strategy achieves rapid surrogate refinement. 
The NRMSE shown in Figure~\ref{fig:2D_straddle_fig2}b reaches values below $0.08$ after 60 iterations for all the examined threshold values.
The intermediate and highest targets achieve the best SSIM (Figure~\ref{fig:2D_straddle_fig2}c) of approximately $0.98$, while the lowest target also exhibits high structural fidelity, with an SSIM around $0.95$.
These results demonstrate that the Straddle acquisition function can identify extended high-efficiency operating regions while maintaining high global surrogate reconstruction accuracy.

\section{Results of 4D optimization}
\label{sec:4D}

While the two-dimensional study provides a controlled benchmark for evaluating physics-informed BO against a direct reference grid scan, the advantages of this approach become particularly relevant in higher-dimensional parameter spaces, where exhaustive scanning becomes computationally impractical.
We therefore extend the optimization to four dimensions by including the driving-laser peak intensity $I$ and pulse duration $\tau$, alongside the gas pressure $p$ and medium length $L$.

These parameters are coupled through both the microscopic harmonic response and macroscopic phase matching, making their simultaneous optimization important for identifying favorable HHG conditions.
For example, retaining 100 sampling points per dimension would increase the number of simulations from $10^4$ in 2D to $10^8$ in 4D.
We therefore assess the predictive accuracy of the BO surrogate models against an independent reference dataset of 1024 HarP simulations, distributed throughout the four-dimensional parameter space using a Sobol quasi-random sequence.
These reference points are excluded from BO training and used to evaluate the global predictive accuracy of the different surrogate models.

To extend the physics-informed approach established in the 2D study, we compare the reference parameterization $(p,L,I,\tau)$ with three progressively modified representations.
First, we adopt the logarithmic product--ratio coordinates, which provided the highest GP surrogate reconstruction accuracy in the 2D optimization, while retaining the original laser parameters $I$ and $\tau$:
\begin{equation}
    \left(\log(p/L),\,\log(pL),\,I,\,\tau\right).
\end{equation}
The second parameterization incorporates the pulse-duration dependence reported in Ref.~\cite{westerberg2025influence}.
For a given driving intensity, there is an optimal pulse duration that maximizes CE, with shorter optimal durations generally associated with higher intensities.
The maximum CE obtained at these optimal durations $\tau_{\mathrm{opt}}$ was found to scale approximately as $\tau_{\mathrm{opt}}^{-1/2}$. 
Although this scaling describes the maximum CE rather than the entire parameter landscape, we also investigate whether the transformation improves the GP reconstruction away from the optimal conditions.
We therefore retain $I$ as an independent coordinate, while replacing $\tau$ with $1/\sqrt{\tau}$ to reflect the observed dependence of the maximum CE on pulse duration:
\begin{equation}\label{eq:transform}
    \left(\log(p/L),\,\log(pL),\,I,\,1/\sqrt{\tau}\right).
\end{equation}
Finally, the third variant combines this parameterization with a square-root transformation of the objective function, $\sqrt{\mathrm{CE}}$, to reduce variations in scale across the CE landscape.

All four optimization strategies employ the Straddle acquisition function with a target conversion efficiency of $\CE_{\mathrm{target}}=3.0\times10^{-6}$, chosen as a representative high-efficiency threshold for exploring favorable operating regions in the four-dimensional parameter space. 
For each parameterization, BO was performed for 120 iterations following an initial evaluation at $i=0$.
Similarly to the 2D study, the GP predictions and reconstruction metrics correspond to observations through iteration $i=119$, whereas the best-so-far CE includes the evaluation at $i=120$.

\begin{figure}[t!]
    \centering
    \includegraphics[width=1.0\linewidth]{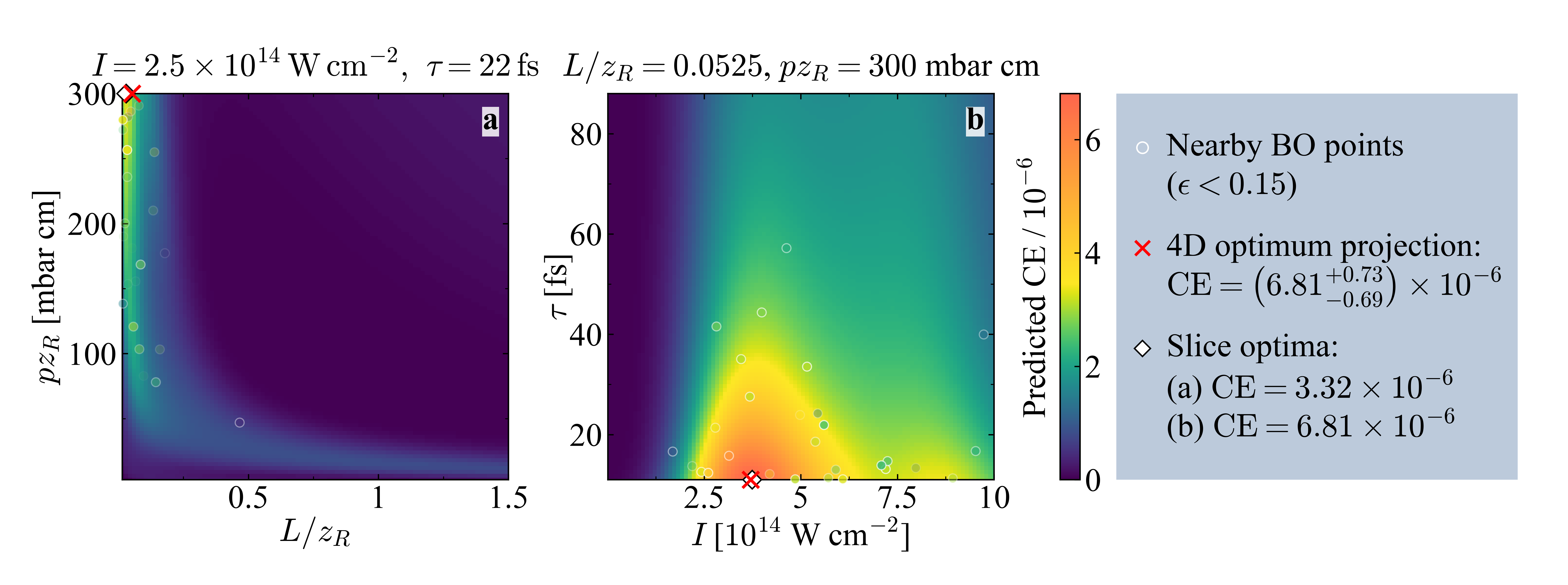}
    \caption{
\textbf{Two-dimensional slices of the CE landscape predicted by the 4D GP surrogate fitted through BO iteration $i=119$.}
    \textbf{a)} Slice in the $L/z_R$ versus $p z_R$ plane at the fixed 2D reference intensity ($I = 2.5 \times 10^{14}~\mathrm{W\,cm^{-2}}$) and pulse duration ($\tau = 22\,\text{fs}$), matching the parameters in 2D study presented in Section \ref{sec:2D}.
     \textbf{b)} Slice in the intensity $I$ versus pulse duration $\tau$ plane evaluated at the optimal 4D spatial parameters ($ L/z_R=0.05$, $p z_R = 300 \,\text{mbar cm}$).
The continuous colormaps represent the GP posterior mean, while the circular markers indicate nearby BO evaluation points, colored by their simulated CE.
Proximity is quantified by the Euclidean distance $\epsilon$ in the two fixed coordinates of each slice, normalized by their respective optimization ranges. Only points with $\epsilon < 0.15$ are shown, with opacity decreasing linearly to zero at the threshold.
The red cross marks the projection of the GP-predicted 4D optimum onto each slice, with a predicted $\mathrm{CE} = (6.81^{+0.73}_{-0.69}) \times 10^{-6}$, where the asymmetric bounds represent the GP's one-standard-deviation uncertainty interval. 
The white diamonds mark the GP-predicted maximum within each slice.
    }
    \label{fig:4Dbo_slices}
\end{figure}

Figure~\ref{fig:4Dbo_slices} illustrates two-dimensional slices of the CE landscape predicted by the four-dimensional GP surrogate using the physics-informed $(\log(p/L),\log(pL),I,1/\sqrt{\tau})$ parameterization with the square-root transformation of CE.
Figure~\ref{fig:4Dbo_slices}a shows the pressure--length landscape at the fixed laser parameters used in the 2D study, revealing the characteristic hyperbolic structure.
Figure~\ref{fig:4Dbo_slices}b shows the intensity--duration dependence at the GP-predicted optimal pressure and medium length, revealing a high-efficiency region at shorter pulse durations and driving intensities around $4\times10^{14}~\mathrm{W\,cm^{-2}}$.

The GP-predicted maximum within the investigated parameter domain occurs at $L/z_R=0.0525$, $pz_R=300\,\mathrm{mbar\,cm}$, $I=3.70\times10^{14}~\mathrm{W\,cm^{-2}}$, and $\tau=11\,\mathrm{fs}$, with a predicted $\CE=(6.81^{+0.73}_{-0.69})\times\nobreak10^{-6}$.
The superscript and subscript indicate the upper and lower one-standard-deviation uncertainty bounds, respectively, which are asymmetric due to the nonlinear transformation from $\sqrt{\CE}$ back to CE.
A confirmatory HarP simulation at these parameters yields $\CE=6.96\times10^{-6}$, in close agreement with the GP prediction.
Notably, the predicted optimum lies at the upper pressure bound, $pz_R=300\,\mathrm{mbar\,cm}$, and the lower pulse-duration bound, $\tau=11\,\mathrm{fs}$. While the present optimization identifies a high-efficiency configuration within the investigated parameter domain, higher conversion efficiencies beyond these limits cannot be excluded. Moreover, the possibility of a different optimal pulse duration in the low-pressure regime could be explored through a separate optimization of this branch.

\begin{figure}[h!]
  \centering
  \includegraphics[width=1\linewidth]{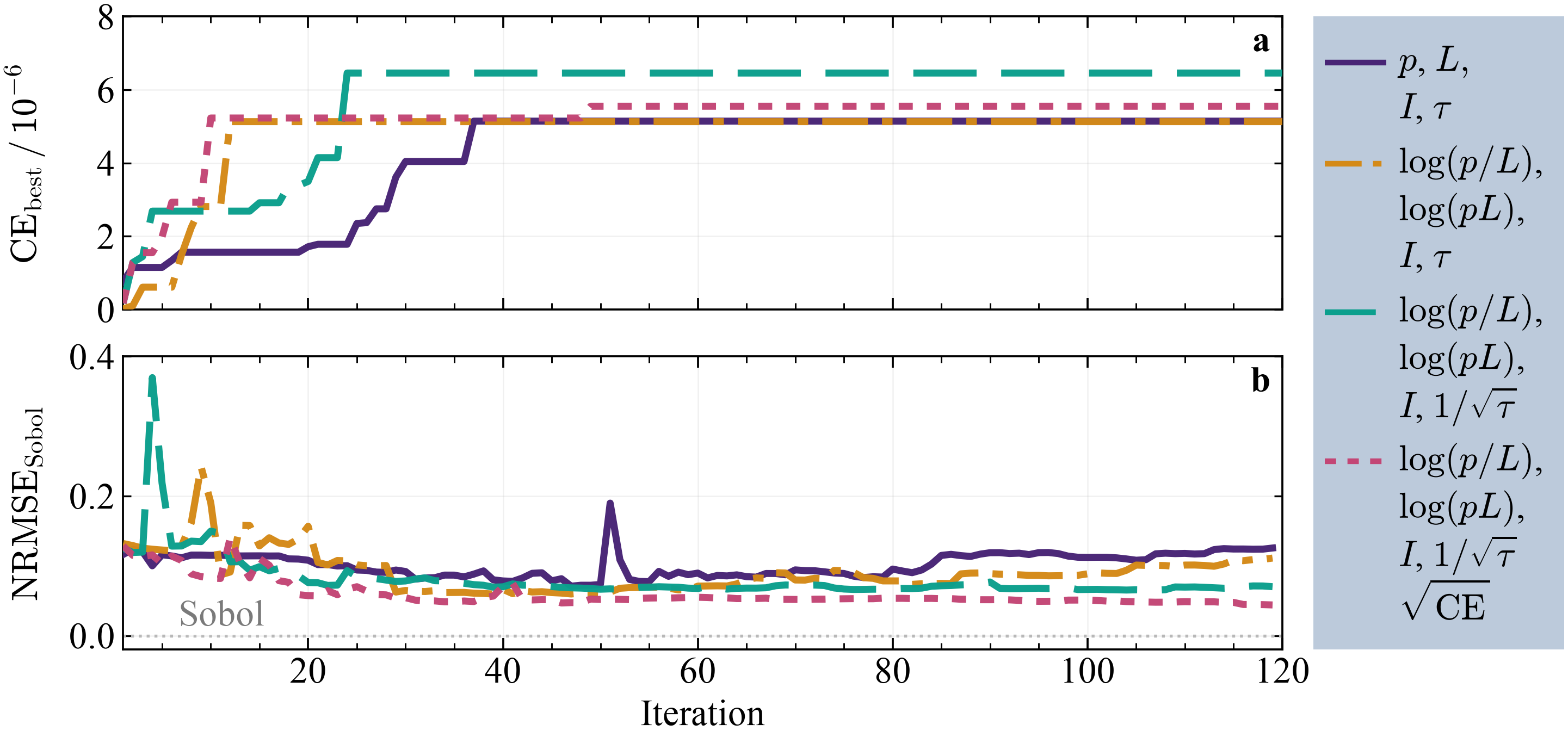}
  \caption{
  \textbf{Convergence and GP prediction accuracy for different 4D BO strategies.}
  \textbf{a)} Evolution of the best-so-far CE as a function of BO iteration.
  \textbf{b)} Normalized root-mean-square error between the GP prediction and the 1024-point Sobol reference dataset $\mathrm{NRMSE}_{\mathrm{Sobol}}$ as a function of BO iteration.
  The dotted gray horizontal line indicates the value of 0, corresponding to perfect agreement with the reference dataset.
  Different colors and marker styles distinguish four different BO parameterizations, all using the Straddle acquisition function with $\CE_{\mathrm{target}}=3.0\times10^{-6}$.
  }
\label{fig:4Dbo_convergence}
\end{figure}

Figure~\ref{fig:4Dbo_convergence}a compares the evolution of the highest sampled CE for the four optimization strategies. 
After the same number of 120 BO iterations, the physics-informed
$(\log(p/L),\log(pL),I,1/\sqrt{\tau})$ 
parameterization achieves the highest sampled CE of approximately $6.46\times10^{-6}$, compared with $5.14\times10^{-6}$ in the reference parameterization $(p,L,I,\tau)$.
The additional square-root transformation of CE yields a somewhat lower sampled maximum of $5.55\times10^{-6}$.
However, since the Straddle acquisition function aims to identify the boundary of a prescribed high-efficiency region rather than locate the global maximum, the highest sampled CE is presented as a complementary measure of optimization performance.

Figure~\ref{fig:4Dbo_convergence}b compares the global predictive accuracy of the GP surrogates against the independent reference dataset of 1024 Sobol-sampled HHG simulations.
Both parameterizations incorporating the inverse square-root pulse-duration scaling achieve better surrogate reconstruction than the parameterization retaining the original pulse duration.
The highest predictive accuracy is achieved by combining the physics-informed coordinates with the square-root transformation of CE, yielding $\mathrm{NRMSE}_{\mathrm{Sobol}}\approx0.044$, compared with $0.126$ for the reference parameterization $(p,L,I,\tau)$, a reduction of approximately $65\%$.
For the same physics-informed coordinates defined in Equation~\eqref{eq:transform}, applying the additional square-root transformation of CE reduces the reconstruction error by approximately $37\%$.

As in the two-dimensional study, the parameterization yielding the highest sampled CE does not necessarily provide the most accurate reconstruction of the underlying landscape. 
These four-dimensional results show that transforming the objective function, in addition to the optimization coordinates, can substantially improve the global predictive accuracy of the GP surrogate for a fixed number of BO iterations.

\section{Conclusions}
\label{sec:conclusions}

We have developed a physics-informed Bayesian optimization framework for efficiently identifying favorable operating conditions and reconstructing multidimensional conversion-efficiency landscapes in gas-based HHG.
By incorporating known physical scaling relations through nonlinear coordinate transformations, the framework improves the accuracy of GP surrogate reconstruction using a limited number of simulations.

We first demonstrated the approach in the 2D pressure--length space, where a direct grid scan provided a reference landscape for evaluating the GP surrogate accuracy.
Using only 60 BO iterations, compared with the $10^4$ simulations required for the direct grid scan, the physics-informed $(\log(pL),\log(p/L))$ parameterization accurately reproduced the characteristic hyperbolic ridge, raising the structural similarity with the reference landscape from 0.77 in the original coordinates to 0.98.
The comparison demonstrates that the improved reconstruction benefits from combining suitable coordinate scaling with alignment to the underlying physical structure of the CE landscape.
Beyond identifying comparable maximum conversion efficiencies, physics-informed coordinates provided a substantially more accurate reconstruction of the surrounding landscape.

This approach was then extended to four dimensions by including the driving-laser intensity and pulse duration, where exhaustive scanning becomes computationally impractical.
Incorporating the inverse square-root pulse-duration scaling and an additional square-root transformation of CE reduced the global surrogate reconstruction error by approximately $65\%$ compared with the reference parameterization $(p,L,I,\tau)$ without the CE transformation, as validated against an independent dataset of 1024 HHG simulations.
The reconstructed 4D landscape revealed a high-efficiency region at short pulse durations and intensities around $4\times10^{14}~\mathrm{W\,cm^{-2}}$, with the predicted optimum lying at the upper pressure and lower pulse-duration bounds of the investigated domain.

Beyond locating favorable operating points, the Straddle acquisition strategy enabled the mapping of extended regions satisfying prescribed CE thresholds.
These high-efficiency regions encompass a range of parameter combinations rather than isolated optimal conditions, providing greater flexibility in the selection of experimentally feasible operating parameters.

The present study also opens several directions for further development. 
In particular, the parameterizations considered here are not necessarily optimal, and alternative transformations could be investigated to further improve surrogate reconstruction accuracy. 
A natural next step is to extend the investigated parameter ranges and increase the dimensionality of the optimization by including additional physical parameters, such as the laser focal position, the gas-density profile, the driving-laser wavelength, or gas species. 
Beyond conversion efficiency, the framework could also be adapted to optimize other HHG observables, provided that suitable physical dependencies can be identified and incorporated into the parameterization. 
More broadly, this approach could be applied to other computationally expensive laser--matter interaction problems where physical knowledge can guide the exploration of multidimensional parameter spaces.
Together, these results establish physics-informed BO as an efficient approach for both optimizing gas-based HHG sources and reconstructing their multidimensional conversion-efficiency landscapes.

\section*{Data availability statement}
The HarP-BO code, the run data and the analysis scripts supporting this study are openly available on GitHub at \url{https://github.com/HarmonicPropagation/HarP-BO} and archived on Zenodo \cite{harp_bo_zenodo}. The larger simulation and optimization datasets are available
separately on Zenodo \cite{harp_bo_data_zenodo}.

\section*{Acknowledgements}
The authors gratefully acknowledge fruitful discussions with Stefanos Carlström, István Pusztai, Melvin Redon, Daniel van Dyk, Anne-Lise Viotti and Christian Russberg.

\section*{Funding}
This project received funding from the Knut and Alice Wallenberg Foundation (Grant Nos. KAW 2020.0111, 2023.0249,  and 2023.0207) and from the Swedish Research Council (Grant No.~2023-04603). A. L. acknowledges support from the Knut and Alice Wallenberg Foundation through the Wallenberg Center for Quantum Technology.
The computations were enabled by resources provided by the National Academic Infrastructure for Supercomputing in Sweden (NAISS), partially funded by the Swedish Research Council through grant agreement No. 2022-06725.


\bibliographystyle{iopart-num}
\bibliography{references}

\end{document}

%% file: bo_scheme.tex

\definecolor{boPurple}{HTML}{440154}
\definecolor{boTeal}{HTML}{21918C}
\definecolor{boPurpleFill}{HTML}{F0ECF5}
\definecolor{boTealFill}{HTML}{E8F5F2}

\begin{tikzpicture}[
    node distance=9mm and 13mm,
    bostep/.style={
        draw=boTeal,
        fill=boTealFill,
        text=black,
        rounded corners=2mm,
        line width=0.8pt,
        minimum width=30mm,
        minimum height=12mm,
        align=center,
        font=\normalsize
    },
    init/.style={
        bostep,
        draw=boPurple,
        fill=boPurpleFill
    },
    flow/.style={
        -{Stealth[length=2.2mm,width=1.7mm]},
        line width=0.8pt,
        draw=boTeal!70!black
    }
]

\node[bostep] (gp) {Update\\GP};

\node[bostep, right=of gp] (acq)
    {Optimise\\acquisition};

\node[bostep, below=of acq] (sim)
    {Run HHG\\simulation};

\node[bostep, below=of gp] (ce)
    {Extract\\CE};

\node[init, left=of gp] (ini)
    {Initial\\CE evaluation};

\draw[
    -{Stealth[length=2.2mm,width=1.7mm]},
    line width=0.8pt,
    draw=boPurple
] (ini.east) -- (gp.west);

\draw[flow]
    (gp.east) to[out=15,in=165] (acq.west);

\draw[flow]
    (acq.south) to[out=-65,in=65] (sim.north);

\draw[flow]
    (sim.west) to[out=195,in=-15] (ce.east);

\draw[flow]
    (ce.north) to[out=115,in=-115] (gp.south);

\node[
    font=\normalsize,
    text=boTeal!70!black,
    align=center
] at ($(gp)!0.5!(sim)$)
    {repeat\\$N_{\mathrm{BO}}$ times};

\end{tikzpicture}

%% file: references.bib
@article{Constant2025perspective,
    author = {Constant, E. and Nandi, S. and Picot, C. and Prost, E. and Palakkal, S. and Lépine, F. and Loriot, V.},
    title = {High order harmonic generation-based attosecond light sources and applications to quantum phenomena},
    journal = {APL Photonics},
    volume = {10},
    number = {1},
    pages = {010907},
    year = {2025},
    month = {01},
    issn = {2378-0967},
    doi = {10.1063/5.0235171},
    url = {https://doi.org/10.1063/5.0235171},
    eprint = {https://pubs.aip.org/aip/app/article-pdf/doi/10.1063/5.0235171/20369740/010907_1_5.0235171.pdf},
}

@article{5D_modelling_roadmap,
	author={Demircan, Ayhan and Evlyukhin, Andrey B. and Morgner, Uwe and Melchert, Oliver and Babushkin, Ihar and Zhang, Shihao and Biancalana, Fabio and Mei, Chao and Fan, Jintao and Yang, Peilong and Steinmeyer, Günter and Lalanne, Philippe and Miller, Owen D and Barati Sedeh, Hooman and Litchinitser, Natalia M and Vaičaitis, Virgilijus and Balachninaitė, Ona and Takayama, Osamu and Malureanu, Radu and Lavrinenko, Andrei V and Wetzel, Benjamin and Chaves, Bruno P and Tonello, Alessandro and Couderc, Vincent and Tuz, Vladimir R and Husakou, Anton and Bergé, Luc and Davoine, Xavier and Gremillet, Laurent and Vábe, Jan and Catoire, Fabrice and Skupin, Stefan and Polynkin, Pavel and Barker, Jacob and Palastro, John and Sergeyev, Sergey and Kbashi, Hani and Turitsyn, Sergei K and Menshikov, Evgenii and Franceschini, Paolo and De Angelis, Costantino and Gorkunov, Maxim V and Kivshar, Yuri S and Beuchler, S. and Endtmayer, B. and Langer, U. and Schafelner, A. and Wick, T. and Binkowski, Felix and Betz, Fridtjof and Hammerschmidt, Martin and Schneider, Philipp-Immanuel and Zschiedrich, Lin and Burger, Sven and Mahlau, Yannik and Berg, Lukas and Lindauer, Marius and Rosenhahn, Bodo and Garg, Puneet and Asadchy, Viktar and Galiffi, Emanuele and Horsley, Simon and Sapienza, Riccardo and Rockstuhl, Carsten and Karabchevsky, Alina and Schrödel, Yannick and Malang, Hanna and Wismath, Sonja and Soennecken, Sören and Helge Dörsam, Jan and Lang, Tino and Kupnik, Mario and Tuennermann, Henrik and Heyl, Christoph M and Stiller, Birgit and Zheng, Lei and Günther, Axel and Kowalsky, Wolfgang and Roth, Bernhard and Duc Tran, Jimmy and Babicheva, Viktoriia E and Chabi, Sakineh and Crandall Jones, Andrew and Kim, Heungsoo and Lock, Evgeniya H and Sun, Kaili and Han, Zhanghua and Hinkelmann, Moritz and Wienke, Andreas and Kues, Michael and Kracht, Dietmar and Calà Lesina, Antonio and Fröhlich, Markus and Biesterfeld, Leon and Klepzig, Lars and Rudolph, Dominik A and Strolka, Onno and Niebur, André and Unseld, Janosch and Kahlmeyer, Leo and Antanovich, Artsiom and Lauth, Jannika and Fütterer, Laura and Overmeyer, Ludger and Schlein, Elena V and Vasylkovskyi, Volodymyr S and Chichkov, Boris N and Wohletz, Jona and Falkner, Malte and Wüllner, Ulf and Biermann, Tobias and Lachmayer, Roland and Rübeling, Philip and Kues, Michael and Malomed, Boris A and Wegener, Lucas and Walla, Peter J and Athanasiou, Stavros and Martin, Olivier J F and Krause, Pascal and Bande, Annika},
	title={Roadmap on 5D Photonics for nano and beyond},
	journal={Nano Futures},
	url={http://iopscience.iop.org/article/10.1088/2399-1984/ae8631},
	year={2026}
}

@misc{attosecond_roadmap,
  doi = {10.48550/ARXIV.2604.15085},
  url = {https://arxiv.org/abs/2604.15085},
  author = {Varillas,  Rocio Borrego and Agostini,  Pierre and Ardana-Lamas,  Fernando and Arnold,  Cord L. and Ayuso,  David and Reduzzi,  Maurizio and Benda,  Jakub and Biegert,  Jens and Bourassin-Bouchet,  Charles and Brabec,  Thomas and Brahms,  Christian and Brown,  Andrew C. and Busto,  David and Caillat,  Jérémie and Calegari,  Francesca and Callegari,  Carlo and Carlstr\"{o}m,  Stefanos and Chang,  Zenghu and Chen,  Ming-Chang and Ciriolo,  Anna G. and Corkum,  Paul and Crippa,  Gabriele and Garcia,  Rafael de Q. and DiMauro,  Louis and Dudovich,  Nirit and Eng-Johnsson,  Per and Faccialà,  Davide and Flores,  Philip and Gadeyne,  Titouan and Geloni,  Gianluca Aldo and Geirger,  Chase and Gholam-Mirzaei,  Shima and Gorfinkiel,  Jimena D. and Goulielmakis,  Eleftherios and Hassan,  Mohammed and Hernández-García,  Carlos and Ho,  Phay and Hui,  Dandan and Hutcheson,  Lynda R. and Ivanov,  Misha and Kahaly,  Subhendu and Kapteyn,  Henry and Karpowicz,  Nicholas and K\"{a}rtner,  Franz X. and Kling,  Matthias and Kneller,  Omer and Ko,  Dong Hyuk and Kraus,  Peter M. and Kubullek,  Maximilian and Leone,  Stephen R. and Lépine,  Franck and L'Huillier,  Anne and Liao,  Chen-Ting and Linker,  Thomas and Lohr,  Alexander Gabriel and Lucchini,  Matteo and Madsen,  Lars Bojer and Mainz,  Roland E. and Major,  Balázs and Marangos,  Jon P. and Marco,  David and Marroux,  Hugo and Marshallsay,  Sean and Vázquez,  Rebeca Martínez and Martín-Hernández,  Rodrigo and Mašín,  Zdeněk and Meyer,  Michael and Moreno,  Felipe Morales and Murnane,  Margaret and Neumark,  Daniel M. and Nisoli,  Mauro and Ossiander,  Marcus and Palakka,  Sreelakshmi and Patchkovskii,  Serguei and Pi,  Zekun and Plaja,  Luis and Poborska,  Julita and Porras,  Miguel A. and Prince,  Kevin C. and Purschke,  David N. and Puskar,  Nicolette G. and Rossi,  Giulio Maria and Rouxel,  Jérémy R. and Ruchon,  Thierry and Rupprecht,  Patrick and Salières,  Pascal and Sansone,  Giuseppe and Scheiba,  Fabian and Schultze,  Martin and Sch\"{u}tte,  Bernd and Serkez,  Svitozar and Silva-Toledo,  Miguel A. and Smirnova,  Olga and Stagira,  Salvatore and Trabattoni,  Andrea and Travers,  John C. and Tyulnev,  Igor and Vacher,  Morgane and Vampa,  Giulio and van der Hart,  Hugo W. and Varjú,  Katalin and Viotti,  Anne-Lise and Vishnoi,  Vartika and Vrakking,  Marc and Wanie,  Vincent and Witte,  Stefan and Xu,  Fei and Yakovlev,  Vladislav S. and Young,  Linda and Zhu,  Diling and Vozzi,  Caterina},
  title = {Roadmap on Attosecond Science},
  publisher = {arXiv},
  year = {2026},
  copyright = {Creative Commons Attribution Non Commercial No Derivatives 4.0 International}
}

@article{lhuillier1992,
  title = {Calculations of high-order harmonic-generation processes in xenon at 1064 nm},
  author = {L'Huillier, Anne and Balcou, Philippe and Candel, Sebastien and Schafer, Kenneth J. and Kulander, Kenneth C.},
  journal = {Physical Review A},
  volume = {46},
  issue = {5},
  pages = {2778--2790},
  numpages = {0},
  year = {1992},
  month = {Sep},
  publisher = {American Physical Society},
  doi = {10.1103/PhysRevA.46.2778},
  url = {https://link.aps.org/doi/10.1103/PhysRevA.46.2778}
}

@article{weissenbilder2022optimize,
  title={How to optimize high-order harmonic generation in gases},
  author={Weissenbilder, R and Carlstr{\"o}m, S and Rego, L and Guo, C and Heyl, CM and Smorenburg, P and Constant, Eric and Arnold, CL and L’Huillier, A},
  journal={Nature Reviews Physics},
  volume={4},
  number={11},
  pages={713--722},
  year={2022},
  publisher={Nature Publishing Group UK London},
  url={https://doi.org/10.1038/s42254-022-00522-7}
}

@article{mcpherson1987studies,
  title={Studies of multiphoton production of vacuum-ultraviolet radiation in the rare gases},
  author={McPherson, A and Gibson, G and Jara, H and Johann, U and Luk, Ting S and McIntyre, IA and Boyer, Keith and Rhodes, Charles K},
  journal={Journal of the Optical Society of America B},
  volume={4},
  number={4},
  pages={595--601},
  year={1987},
  publisher={Optical Society of America},
  url={https://doi.org/10.1364/JOSAB.4.000595}
}

@article{ferray1988multiple,
  title={Multiple-harmonic conversion of 1064 nm radiation in rare gases},
  author={Ferray, M and L'Huillier, Anne and Li, XF and Lompre, LA and Mainfray, G and Manus, C},
  journal={Journal of Physics B: Atomic, Molecular and Optical Physics},
  volume={21},
  number={3},
  pages={L31--L35},
  year={1988},
  url ={https://doi.org/10.1088/0953-4075/21/3/001}
}

@misc{vabek2026modularmultiscaleapproachmodelling,
      title={Modular multiscale approach to modelling high-harmonic generation in gases}, 
      author={Jan Vábek and Tadeáš Němec and Stefan Skupin and Fabrice Catoire},
      year={2026},
      eprint={2507.04115},
      archivePrefix={arXiv},
      primaryClass={physics.optics},
      url={https://arxiv.org/abs/2507.04115}, 
}

@article{Constant1999,
  title = {Optimizing High Harmonic Generation in Absorbing Gases: Model and Experiment},
  author = {Constant, E. and Garzella, D. and Breger, P. and M\'evel, E. and Dorrer, Ch. and Le Blanc, C. and Salin, F. and Agostini, P.},
  journal = {Physical Review Letters},
  volume = {82},
  issue = {8},
  pages = {1668--1671},
  numpages = {0},
  year = {1999},
  month = {Feb},
  publisher = {American Physical Society},
  doi = {10.1103/PhysRevLett.82.1668},
  url = {https://link.aps.org/doi/10.1103/PhysRevLett.82.1668}
}

@article{LHuillier1996,
  title = {Theory of high-order harmonic generation by an elliptically polarized laser field},
  author = {Antoine, Philippe and L'Huillier, Anne and Lewenstein, Maciej and Sali\`eres, Pascal and Carr\'e, Bertrand},
  journal = {Physical Review A},
  volume = {53},
  issue = {3},
  pages = {1725--1745},
  numpages = {0},
  year = {1996},
  month = {Mar},
  publisher = {American Physical Society},
  doi = {10.1103/PhysRevA.53.1725},
  url = {https://link.aps.org/doi/10.1103/PhysRevA.53.1725}
}

@book{matern1960spatial,
  title={Spatial Variation},
  author={Mat{\'e}rn, Bertil},
  year={1960},
  publisher={Meddelanden fran Statens Skogsforskningsinstitut}
}

@article{sobol1967distribution,
  author  = {Sobol', Ilya M.},
  title   = {On the Distribution of Points in a Cube and the Approximate Evaluation of Integrals},
  journal = {USSR Computational Mathematics and Mathematical Physics},
  volume  = {7},
  number  = {4},
  pages   = {86--112},
  year    = {1967},
  doi     = {10.1016/0041-5553(67)90144-9},
  url     = {https://doi.org/10.1016/0041-5553(67)90144-9}
}

@phdthesis{neal1996bayesian,
  author={Neal, Radford M.},
  title={Bayesian Learning for Neural Networks},
  school={University of Toronto},
  year={1995}
}

@inproceedings{balandat2020botorch,
  title={{BoTorch: A Framework for Efficient Monte-Carlo Bayesian Optimization}},
  author={Balandat, Maximilian and Karrer, Brian and Jiang, Daniel and Daulton, Samuel and Letham, Benjamin and Wilson, Andrew Gordon and Bakshy, Eytan},
  booktitle={Advances in Neural Information Processing Systems},
  volume={33},
  pages={21524--21538},
  year={2020}
}

@inproceedings{gardner2018gpytorch,
  title={{GPyTorch: Blackbox Matrix-Matrix Gaussian Process Inference with GPU Acceleration}},
  author={Gardner, Jacob and Pleiss, Geoff and Weinberger, Kilian and Bindel, David and Wilson, Andrew Gordon},
  booktitle={NeurIPS},
  year={2018}
}

@article{wang2004image,
  title={Image quality assessment: from error visibility to structural similarity},
  author={Wang, Zhou and Bovik, Alan C and Sheikh, Hamid R and Simoncelli, Eero P},
  journal={IEEE Transactions On Image Processing},
  volume={13},
  number={4},
  pages={600--612},
  year={2004},
  publisher={IEEE}
}

@book{rasmussen2006gaussian,
  title={Gaussian Processes for Machine Learning},
  author={Rasmussen, Carl Edward and Williams, Christopher K. I.},
  year={2006},
  publisher={MIT Press},
  url={https://doi.org/10.7551/mitpress/3206.001.0001}
}

@article{jones1998efficient,
  title={Efficient global optimization of expensive black-box functions},
  author={Jones, Donald R and Schonlau, Matthias and Welch, William J},
  journal={Journal of Global Optimization},
  volume={13},
  number={4},
  pages={455--492},
  year={1998},
  publisher={Springer}
}

@article{ament2023unexpected,
  title={Unexpected improvements to expected improvement for {Bayesian} optimization},
  author={Ament, Sebastian and Daulton, Samuel and Eriksson, David and Balandat, Maximilian and Bakshy, Eytan},
  journal={Advances in Neural Information Processing Systems},
  volume={36},
  pages={20577--20612},
  year={2023}
}

@article{soldan2001static,
  title={Static dipole polarizabilities ($\alpha$) and static second hyperpolarizabilities ($\gamma$) of the rare gas atoms (He--Rn)},
  author={Sold{\'a}n, Pavel and Lee, Edmond PF and Wright, Timothy G},
  journal={Physical Chemistry Chemical Physics},
  volume={3},
  number={21},
  pages={4661--4666},
  year={2001},
  publisher={Royal Society of Chemistry},
  url={https://doi.org/10.1039/b105433n
}
}

@article{Cabasse12,
author = {Am\'{e}lie Cabasse and Guillaume Machinet and Antoine Dubrouil and Eric Cormier and Eric Constant},
journal = {Optics Letters},
number = {22},
pages = {4618--4620},
publisher = {Optica Publishing Group},
title = {Optimization and phase matching of fiber-laser-driven high-order harmonic generation at high repetition rate},
volume = {37},
month = {Nov},
year = {2012},
url = {https://opg.optica.org/ol/abstract.cfm?URI=ol-37-22-4618},
doi = {10.1364/OL.37.004618},
}

@article{Heyl_2017,
doi = {10.1088/1361-6455/50/1/013001},
url = {https://doi.org/10.1088/1361-6455/50/1/013001},
year = {2016},
month = {dec},
publisher = {IOP Publishing},
volume = {50},
number = {1},
pages = {013001},
author = {Heyl, C M and Arnold, C L and Couairon, A and L’Huillier, A},
title = {Introduction to macroscopic power scaling principles for high-order harmonic generation},
journal = {Journal of Physics B: Atomic, Molecular and Optical Physics}
}

@article{nibbering1997determination,
  title={Determination of the inertial contribution to the nonlinear refractive index of air, N2, and O2 by use of unfocused high-intensity femtosecond laser pulses},
  author={Nibbering, ETJ and Grillon, G and Franco, Michel A and Prade, BS and Mysyrowicz, Andr{\'e}},
  journal={Journal of the Optical Society of America B},
  volume={14},
  number={3},
  pages={650--660},
  year={1997},
  publisher={Optical Society of America},
  url={https://doi.org/10.1364/JOSAB.14.000650}
}

@article{BelmiroChu:21,
author = {Chon-Teng Belmiro Chu and Yae-Lin Sheu and Shih-I Chu},
journal = {Optics Express},
number = {21},
pages = {32900--32909},
publisher = {Optica Publishing Group},
title = {Bayesian optimal control of the ultrashort circularly polarized attosecond pulse generation by two-color polarization gating},
volume = {29},
month = {Oct},
year = {2021},
url = {https://opg.optica.org/oe/abstract.cfm?URI=oe-29-21-32900},
doi = {10.1364/OE.438212},
}

@ARTICLE{BOreview,
  author={Shahriari, Bobak and Swersky, Kevin and Wang, Ziyu and Adams, Ryan P. and de Freitas, Nando},
  journal={Proceedings of the IEEE}, 
  title={Taking the Human Out of the Loop: A Review of Bayesian Optimization}, 
  year={2016},
  volume={104},
  number={1},
  pages={148-175},
  doi={10.1109/JPROC.2015.2494218},
  url={https://doi.org/10.1109/JPROC.2015.2494218}
  }

@article{KrauszIvanov,
  title = {Attosecond physics},
  author = {Krausz, Ferenc and Ivanov, Misha},
  journal = {Reviews of Modern Physics},
  volume = {81},
  issue = {1},
  pages = {163--234},
  numpages = {0},
  year = {2009},
  month = {Feb},
  publisher = {American Physical Society},
  doi = {10.1103/RevModPhys.81.163},
  url = {https://link.aps.org/doi/10.1103/RevModPhys.81.163}
}

@article{kerr1875xl,
  title={XL. A new relation between electricity and light: Dielectrified media birefringent},
  author={Kerr, John},
  journal={The London, Edinburgh, and Dublin Philosophical Magazine and Journal of Science},
  volume={50},
  number={332},
  pages={337--348},
  year={1875},
  publisher={Taylor \& Francis},
  url={https://doi.org/10.1080/14786447508641302}
}

@article{crank1947practical,
  author  = {Crank, J. and Nicolson, P.},
  title   = {A Practical Method for Numerical Evaluation of Solutions of Partial Differential Equations of the Heat-Conduction Type},
  journal = {Proceedings of the Cambridge Philosophical Society},
  volume  = {43},
  number  = {1},
  pages   = {50--67},
  year    = {1947},
  doi     = {10.1017/S0305004100023197},
  url     = {https://doi.org/10.1017/S0305004100023197 }
}

@article{henke1993x,
  title={X-ray interactions: photoabsorption, scattering, transmission, and reflection at E= 50-30,000 eV, Z= 1-92},
  author={Henke, Burton L and Gullikson, Eric M and Davis, John C},
  journal={Atomic data and nuclear data tables},
  volume={54},
  number={2},
  pages={181--342},
  year={1993},
  publisher={Elsevier},
  url={https://doi.org/10.1006/adnd.1993.1013}
}

@article{Corkum,
  title = {Plasma perspective on strong field multiphoton ionization},
  author = {Corkum, P. B.},
  journal = {Physical Review Letters},
  volume = {71},
  issue = {13},
  pages = {1994--1997},
  numpages = {0},
  year = {1993},
  month = {Sep},
  publisher = {American Physical Society},
  doi = {10.1103/PhysRevLett.71.1994},
  url = {https://link.aps.org/doi/10.1103/PhysRevLett.71.1994}
}

@article{Lewenstein,
  title = {Theory of high-harmonic generation by low-frequency laser fields},
  author = {Lewenstein, M. and Balcou, Ph. and Ivanov, M. Yu. and L'Huillier, Anne and Corkum, P. B.},
  journal = {Physical Review A},
  volume = {49},
  issue = {3},
  pages = {2117--2132},
  numpages = {0},
  year = {1994},
  month = {Mar},
  publisher = {American Physical Society},
  doi = {10.1103/PhysRevA.49.2117},
  url = {https://link.aps.org/doi/10.1103/PhysRevA.49.2117}
}

@article{Appi:23,
author = {E. Appi and R. Weissenbilder and B. Nagyill\'{e}s and Z. Diveki and J. Peschel and B. Farkas and M. Plach and F. Vismarra and V. Poulain and N. Weber and C. L. Arnold and K. Varj\'{u} and S. Kahaly and P. Eng-Johnsson and A. L'Huillier},
journal = {Optics Express},
number = {20},
pages = {31687--31697},
publisher = {Optica Publishing Group},
title = {Two phase-matching regimes in high-order harmonic generation},
volume = {31},
month = {Sep},
year = {2023},
url = {https://opg.optica.org/oe/abstract.cfm?URI=oe-31-20-31687},
doi = {10.1364/OE.488298},
}

@article{zurch2014real,
  title={Real-time and sub-wavelength ultrafast coherent diffraction imaging in the extreme ultraviolet},
  author={Z{\"u}rch, M and Rothhardt, J and H{\"a}drich, S and Demmler, S and Krebs, M and Limpert, Jens and T{\"u}nnermann, Andreas and Guggenmos, Alexander and Kleineberg, Ulf and Spielmann, C},
  journal={Scientific Reports},
  volume={4},
  number={1},
  pages={7356},
  year={2014},
  publisher={Nature Publishing Group UK London},
  url = {https://doi.org/10.1038/srep07356}
}

@article{sheehy1999high,
  title={High harmonic generation at long wavelengths},
  author={Sheehy, B and Martin, JDD and DiMauro, LF and Agostini, P and Schafer, KJ and Gaarde, MB and Kulander, KC},
  journal={Physical Review Letters},
  volume={83},
  number={25},
  pages={5270},
  year={1999},
  publisher={APS},
  url ={https://doi.org/10.1103/PhysRevLett.83.5270}
}

@article{brichta2009comparison,
  title={Comparison and real-time monitoring of high-order harmonic generation in different sources},
  author={Brichta, J-P and Wong, MCH and Bertrand, JB and Bandulet, H-C and Rayner, DM and Bhardwaj, VR},
  journal={Physical Review A—Atomic, Molecular, and Optical Physics},
  volume={79},
  number={3},
  pages={033404},
  year={2009},
  publisher={APS},
  url={https://doi.org/10.1103/PhysRevA.79.033404}
}

@article{tempea2000optimization,
  title={Optimization of high-harmonic generation},
  author={Tempea, G and Brabec, T},
  journal={Applied Physics B},
  volume={70},
  number={Suppl 1},
  pages={S197--S202},
  year={2000},
  publisher={Springer},
  url={https://doi.org/10.1007/s003400000310}
}

@article{mikaelsson2020high,
  title={A high-repetition rate attosecond light source for time-resolved coincidence spectroscopy},
  author={Mikaelsson, Sara and Vogelsang, Jan and Guo, Chen and Sytcevich, Ivan and Viotti, Anne-Lise and Langer, Fabian and Cheng, Yu-Chen and Nandi, Saikat and Jin, Wenjie and Olofsson, Anna and others},
  journal={Nanophotonics},
  volume={10},
  number={1},
  pages={117--128},
  year={2020},
  publisher={De Gruyter},
  url={https://doi.org/10.1515/nanoph-2020-0424}
}

@article{westerberg2025influence,
  title={Influence of the laser pulse duration in high-order harmonic generation},
  author={Westerberg, Saga and Redon, Melvin and Raab, A-K and Beaufort, Gaspard and Arias Velasco, Marta and Guo, Chen and Sytcevich, Ivan and Weissenbilder, Robin and O’Dwyer, David and Smorenburg, Peter and others},
  journal={APL Photonics},
  volume={10},
  number={9},
  year={2025},
  publisher={AIP Publishing},
  url={https://doi.org/10.1063/5.0272968
}
}

@article{popmintchev2012bright,
  title={Bright coherent ultrahigh harmonics in the keV x-ray regime from mid-infrared femtosecond lasers},
  author={Popmintchev, Tenio and Chen, Ming-Chang and Popmintchev, Dimitar and Arpin, Paul and Brown, Susannah and Ali{\v{s}}auskas, Skirmantas and Andriukaitis, Giedrius and Bal{\v{c}}iunas, Tadas and M{\"u}cke, Oliver D and Pugzlys, Audrius and others},
  journal={Science},
  volume={336},
  number={6086},
  pages={1287--1291},
  year={2012},
  publisher={American Association for the Advancement of Science},
  url={https://doi.org/10.1126/science.1218497}
}

@article{brabec2000intense,
  title={Intense few-cycle laser fields: Frontiers of nonlinear optics},
  author={Brabec, Thomas and Krausz, Ferenc},
  journal={Reviews of Modern Physics},
  volume={72},
  number={2},
  pages={545},
  year={2000},
  publisher={APS},
  url ={https://doi.org/10.1103/RevModPhys.72.545}
}

@article{carlstrom2022generala,
  title={General time-dependent configuration-interaction singles. I. Molecular case},
  author={Carlstr{\"o}m, Stefanos and Spanner, Michael and Patchkovskii, Serguei},
  journal={Physical Review A},
  volume={106},
  number={4},
  pages={043104},
  year={2022},
  publisher={APS},
  url={https://doi.org/10.1103/PhysRevA.106.043104}
}

@article{agostini2004physics,
  title={The physics of attosecond light pulses},
  author={Agostini, Pierre and DiMauro, Louis F},
  journal={Reports on Progress in Physics},
  volume={67},
  number={6},
  pages={813--855},
  year={2004},
  url={https://doi.org/10.1088/0034-4885/67/6/R01}
}

@article{carlstrom2022generalb,
  title={General time-dependent configuration-interaction singles. II. Atomic case},
  author={Carlstr{\"o}m, Stefanos and Bertolino, Mattias and Dahlstr{\"o}m, Jan Marcus and Patchkovskii, Serguei},
  journal={Physical Review A},
  volume={106},
  number={4},
  pages={042806},
  year={2022},
  publisher={APS},
  url={https://doi.org/10.1103/PhysRevA.106.042806}
}

@article{gaarde2008macroscopic,
  title={Macroscopic aspects of attosecond pulse generation},
  author={Gaarde, Mette B and Tate, Jennifer L and Schafer, Kenneth J},
  journal={Journal of Physics B: Atomic, Molecular and Optical Physics},
  volume={41},
  number={13},
  pages={132001},
  year={2008},
  url={https://doi.org/10.1088/0953-4075/41/13/132001}
}

@article{popmintchev2018near,
  title={Near-and extended-edge x-ray-absorption fine-structure spectroscopy using ultrafast coherent high-order harmonic supercontinua},
  author={Popmintchev, Dimitar and Galloway, Benjamin R and Chen, Ming-Chang and Dollar, Franklin and Mancuso, Christopher A and Hankla, Amelia and Miaja-Avila, Luis and O’Neil, Galen and Shaw, Justin M and Fan, Guangyu and others},
  journal={Physical Review Letters},
  volume={120},
  number={9},
  pages={093002},
  year={2018},
  publisher={APS},
  url={ https://doi.org/10.1103/PhysRevLett.120.093002}
}

@article{bergues2018tabletop,
  title={Tabletop nonlinear optics in the 100-eV spectral region},
  author={Bergues, Boris and Rivas, DE and Weidman, Matthew and Muschet, AA and Helml, Wolfram and Guggenmos, Alexander and Pervak, V and Kleineberg, Ulf and Marcus, G and Kienberger, Reinhard and others},
  journal={Optica},
  volume={5},
  number={3},
  pages={237--242},
  year={2018},
  publisher={Optical Society of America},
  url={https://doi.org/10.1364/OPTICA.5.000237}
}

@article{nayak2018multiple,
  title={Multiple ionization of argon via multi-XUV-photon absorption induced by 20-GW high-order harmonic laser pulses},
  author={Nayak, Arjun and Orfanos, I and Makos, I and Dumergue, Mathieu and K{\"u}hn, Sergei and Skantzakis, E and Bodi, Bal{\'a}zs and Varj{\'u}, Katalin and Kalpouzos, C and Banks, HIB and others},
  journal={Physical Review A},
  volume={98},
  number={2},
  pages={023426},
  year={2018},
  publisher={APS},
  url={https://doi.org/10.1103/PhysRevA.98.023426}
}

@article{takahashi2013attosecond,
  title={Attosecond nonlinear optics using gigawatt-scale isolated attosecond pulses},
  author={Takahashi, Eiji J and Lan, Pengfei and M{\"u}cke, Oliver D and Nabekawa, Yasuo and Midorikawa, Katsumi},
  journal={Nature Communications},
  volume={4},
  number={1},
  pages={2691},
  year={2013},
  publisher={Nature Publishing Group UK London},
  url={https://doi.org/10.1038/ncomms3691}
}

@article{spears2025predicting,
  title={Predicting fusion ignition at the National Ignition Facility with physics-informed deep learning},
  author={Spears, Brian K and Brandon, Scott and Casey, Dan T and Field, John E and Gaffney, Jim A and Humbird, Kelli D and Kritcher, Andrea L and Kruse, Michael KG and Kur, Eugene and Kustowski, Bogdan and others},
  journal={Science},
  volume={389},
  number={6761},
  pages={727--731},
  year={2025},
  publisher={American Association for the Advancement of Science},
  url={https://doi.org/10.1126/science.adm8201}
}

@article{valenta2025bayesian,
  title={Bayesian optimization of electron energy from laser wakefield accelerators},
  author={Valenta, P and Esirkepov, T Zh and Ludwig, JD and Wilks, SC and Bulanov, SV},
  journal={Physical Review Accelerators and Beams},
  volume={28},
  number={9},
  pages={094601},
  year={2025},
  publisher={APS},
  url={https://doi.org/10.1103/knh7-hbr3}
}

@article{arran2026bayesian,
  title={Bayesian optimisation of non-linear Breit--Wheeler pair production in simulated laser experiments},
  author={Arran, Christopher and Morris, Stuart and Ridgers, Christopher P},
  journal={New Journal of Physics},
  volume={28},
  number={4},
  pages={044304},
  year={2026},
  publisher={IOP Publishing},
  url={https://doi.org/10.1088/1367-2630/ae57cb}
}

@article{shalloo2020automation,
  title={Automation and control of laser wakefield accelerators using Bayesian optimization},
  author={Shalloo, RJ and Dann, SJD and Gruse, J-N and Underwood, CID and Antoine, Andre F and Arran, Christopher and Backhouse, Michael and Baird, CD and Balcazar, MD and Bourgeois, Nicholas and others},
  journal={Nature Communications},
  volume={11},
  number={1},
  pages={6355},
  year={2020},
  publisher={Nature Publishing Group UK London},
  url={https://doi.org/10.1038/s41467-020-20245-6}
}

@article{maslarova2026batch,
  title={Batch Bayesian optimization of attosecond betatron pulses from laser wakefield acceleration},
  author={Maslarova, Dominika and Hansson, Albert and Luo, Mufei and Horn{\'y}, Vojt{\v{e}}ch and Ferri, Julien and Pusztai, Istvan and F{\"u}l{\"o}p, T{\"u}nde},
  journal={Communications Physics},
  volume={9},
  number={1},
  pages={92},
  year={2026},
  publisher={Nature Publishing Group UK London},
  url={https://doi.org/10.1038/s42005-026-02542-6}
}

@article{jalas2021bayesian,
  title={Bayesian optimization of a laser-plasma accelerator},
  author={Jalas, S{\"o}ren and Kirchen, Manuel and Messner, Philipp and Winkler, Paul and H{\"u}bner, Lars and Dirkwinkel, Julian and Schnepp, Matthias and Lehe, Remi and Maier, Andreas R},
  journal={Physical Review Letters},
  volume={126},
  number={10},
  pages={104801},
  year={2021},
  publisher={APS},
  url={https://doi.org/10.1103/PhysRevLett.126.104801}
}

@article{fan2025improved,
  title={Improved Bayesian optimization of laser-driven plasma surface high harmonic generation},
  author={Fan, Lili and Wang, Ziwei and Liao, Chenfei and Wang, Jingwei},
  journal={Optics Express},
  volume={33},
  number={5},
  pages={11254--11263},
  year={2025},
  publisher={Optica Publishing Group},
  url ={https://doi.org/10.1364/OE.549712}
}

@article{YE2022106116,
title = {Fast optimization for betatron radiation from laser wakefield acceleration based on Bayesian optimization},
journal = {Results in Physics},
volume = {43},
pages = {106116},
year = {2022},
issn = {2211-3797},
doi = {https://doi.org/10.1016/j.rinp.2022.106116},
url = {https://www.sciencedirect.com/science/article/pii/S2211379722007306},
author = {Hansheng Ye and Yuqiu Gu and Xiaohui Zhang and Shaoyi Wang and Fang Tan and Jie Zhang and Yue Yang and Yonghong Yan and Yuchi Wu and Wenhui Huang and Weimin Zhou}
}

@article{goulielmakis2010real,
  title={Real-time observation of valence electron motion},
  author={Goulielmakis, Eleftherios and Loh, Zhi-Heng and Wirth, Adrian and Santra, Robin and Rohringer, Nina and Yakovlev, Vladislav S and Zherebtsov, Sergey and Pfeifer, Thomas and Azzeer, Abdallah M and Kling, Matthias F and others},
  journal={Nature},
  volume={466},
  number={7307},
  pages={739--743},
  year={2010},
  publisher={Nature Publishing Group UK London},
  url={https://doi.org/10.1038/nature09212}
}

@article{paul2001observation,
  title={Observation of a train of attosecond pulses from high harmonic generation},
  author={Paul, Pierre-Marie and Toma, Elena S and Breger, Pierre and Mullot, Genevive and Aug{\'e}, Fr{\'e}d{\'e}rika and Balcou, Ph and Muller, Harm Geert and Agostini, Pierre},
  journal={Science},
  volume={292},
  number={5522},
  pages={1689--1692},
  year={2001},
  publisher={American Association for the Advancement of Science},
  url={https://doi.org/10.1126/science.1059413}
}

@incollection{frazier2018tutorial,
  author    = {Frazier, Peter I.},
  title     = {A Tutorial on Bayesian Optimization},
  booktitle = {Recent Advances in Optimization and Modeling of Contemporary Problems},
  series    = {INFORMS Tutorials in Operations Research},
  pages     = {255--278},
  year      = {2018},
  publisher = {INFORMS},
  doi       = {10.1287/educ.2018.0188},
  url       = {https://doi.org/10.1287/educ.2018.0192.contrib}
}

@misc{harmonicpropagation,
  author = {},
  title = {{HarmonicPropagation}: {GitHub} Organization},
  year = {2026},
  howpublished = {\url{https://github.com/HarmonicPropagation}},
  note = {{GitHub} organization}
}

@misc{harp_2026,
  author = {Pouyez, Mattys and Maslarova, Dominika
            and Hagedorn, Linus and Jansson, Patrik
            and Weissenbilder, Robin and Carlstr{\"o}m, Stefanos
            and L'Huillier, Anne and F{\"u}l{\"o}p, T{\"u}nde},
  title = {{HarP}: High-order Harmonic Propagation Simulation Software},
  year = {2026},
  howpublished = {\url{https://github.com/HarmonicPropagation/HarP}},
  note = {{GitHub} repository}
}

@article{Gaiser2018,
  author  = {Gaiser, Christof and Fellmuth, Bernd},
  title   = {Polarizability of Helium, Neon, and Argon: New Perspectives for Gas Metrology},
  journal = {Physical Review Letters},
  volume  = {120},
  number  = {12},
  pages   = {123203},
  year    = {2018},
  doi     = {10.1103/PhysRevLett.120.123203}
}

@article{Polyanskiy2024,
  author  = {Polyanskiy, Mikhail N.},
  title   = {{Refractiveindex.info} database of optical constants},
  journal = {Scientific Data},
  volume  = {11},
  pages   = {94},
  year    = {2024},
  doi     = {10.1038/s41597-023-02898-2}
}

@misc{harp_bo_zenodo,
  title  = {{HarP-BO: Bayesian optimization of high-harmonic generation with HarP},},
  author = {Maslarova, Dominika and Jansson, Patrik and Pouyez, Mattys},
  year   = {2026},
  howpublished = {Zenodo, version v1.0.0},
  note   = {\url{https://doi.org/10.5281/zenodo.24023612}, code at \url{https://github.com/HarmonicPropagation/HarP-BO}}
}

@misc{harp_bo_data_zenodo,
title = {{HarP-BO: Data for Bayesian optimization of high-harmonic generation with HarP},},
  author = {Maslarova, Dominika and Jansson, Patrik
            and Pouyez, Mattys},
  year   = {2026},
  howpublished = {Zenodo, dataset, version v1.0.0},
  note   = {\url{https://doi.org/10.5281/zenodo.24044917}}
}

@misc{Mattyspaper,
  title={{HarP}: a simulation tool for high-order harmonic generation and propagation in gases},
  author={Pouyez, Mattys and Maslarova, Dominika and Hagedorn, Linus and Jansson, Patrik
  and Weissenbilder, Robin and Carlström, Stefanos and L'Huillier, Anne and F\"ul\"op, T\"unde},
  year={2026},
  eprint={2610.35440},
  archivePrefix={arXiv},
  primaryClass={physics.optics},
  url={https://arxiv.org/abs/2610.35440},
}
